\documentclass{aa}             %activate for journal version 

\input epsf

\begin{document}

\def\BK{BK\,Peg}    
\def\kms{\ifmmode{\rm km\thinspace s^{-1}}\else km\thinspace s$^{-1}$\fi}
\def\cm{\ifmmode{\rm cm\thinspace^{-1}}\else cm\thinspace$^{-1}$\fi}
\def\Msun{$M_{\sun}$}
\def\Rsun{$R_{\sun}$}
\def\Lsun{$L_{\sun}$}
\def\feh{$[\mathrm{Fe/H}]$}
\def\meh{$[\mathrm{M/H}]$}
\def\afe{$[\mathrm{\alpha/Fe}]$}
\def\ione{\,{\sc i}}
\def\itwo{\,{\sc ii}}

\title{
Absolute dimensions of eclipsing binaries. XXVIII.
\thanks{Based on observations carried out at the Str{\"o}mgren Automatic 
Telescope (SAT) and the 1.5m telescope (63.H-0080) at ESO, La Silla, 
and the Nordic Optical Telescope at La Palma
}}
\subtitle{BK\,Pegasi and other F-type binaries: \\
Prospects for calibration of convective core overshoot
\thanks{
Tables~13--17 are available in electronic form
at the CDS via anonymous ftp to cdsarc.u-strasbg.fr (130.79.128.5)
or via http://cdsweb.u-strasbg.fr/cgi-bin/qcat?J/A+A/}
}
\author{
J.V. Clausen     \inst{1}
\and
S. Frandsen      \inst{2}
\and
H. Bruntt        \inst{3,4}
\and
E.~H. Olsen      \inst{1}
\and
B.~E. Helt       \inst{1}
\and
K. Gregersen     \inst{1}
\and
D. Juncher       \inst{1}
\and
P. Krogstrup     \inst{1}
}
\offprints{J.V.~Clausen, \\ e-mail: jvc@nbi.ku.dk}

\institute{
Niels Bohr Institute, Copenhagen University,
Juliane Maries Vej 30,
DK-2100 Copenhagen {\O}, Denmark
\and
Department of Physics and Astronomy, University of Aarhus,
Ny Munkegade, DK-8000 Aarhus C, Denmark
\and
Observatoire de Paris, LESIA, 5 Place Jules Janssen,
95195 Meudon, France
\and
Sydney Institute for Astronomy, School of Physics, University of Sydney,
NSW 2006, Australia
}

\date{Received 16 February 2010 / Accepted 26 March 2010}

\titlerunning{BK\,Peg}
\authorrunning{J.V. Clausen et al.}

\abstract
%context {optional}
{
Double-lined, detached eclipsing binaries are our main source for accurate
stellar masses and radii.
In this paper we focus on the 1.15--1.70 \Msun\ interval where convective
core overshoot is gradually ramped up in theoretical evolutionary models.
}
%aims
{We aim to determine absolute dimensions and abundances for the
F-type detached eclipsing binary \BK, and to perform a detailed
comparison with results from recent stellar evo\-lu\-tio\-nary models, 
including a sample of previously studied systems with accurate parameters.} 
%methods
{$uvby$ light curves and $uvby\beta$ standard photometry were obtained with
the Str\"omgren Automatic Telescope, ESO, La Silla, and high-resolution spectra
were acquired with the FIES spectrograph at the Nordic Optical Telescope, La Palma.}
%results
{
The $5\fd49$ period orbit of \BK\ is slightly eccentric ($e$ = 0.053).
The two components are quite different with masses and radii of 
($1.414 \pm 0.007$ \Msun, $1.988 \pm 0.008$ \Rsun) and
($1.257 \pm 0.005$ \Msun, $1.474 \pm 0.017$ \Rsun), respectively.
The measured rotational velocities are $16.6 \pm 0.2$ (primary) and
$13.4 \pm 0.2$ (secondary) \kms. For the secondary component this corresponds
to (pseudo)synchronous rotation, whereas the primary component seems to
rotate at a slightly lower rate. 
We derive an iron abundance of \feh\,$=-0.12\pm0.07$ and similar abundances for 
Si, Ca, Sc, Ti, Cr and Ni.
The stars have evolved to the upper half of the main-sequence band.
Yonsei-Yale and Victoria-Regina evolutionary models for the
observed metal abundance reproduce \BK\ at ages of 2.75 and 2.50 Gyr, respectively, but tend to
predict a lower age for the more massive primary component than for the secondary.
We find the same age trend for three other upper main-sequence systems in
a sample of well studied eclipsing binaries with components in the
1.15--1.70 \Msun\ range.
We also find that the Yonsei-Yale models systematically predict higher ages than the
Victoria-Regina models. 
The sample includes BW\,Aqr, and as a supplement we have determined a
\feh\ abundance of $-0.07 \pm 0.11$ for this late F-type binary.
}
%conclusions {optional}
{We propose to use \BK, BW\,Aqr, and other well-studied 1.15--1.70 \Msun\ eclipsing binaries 
to fine-tune convective core overshoot, diffusion, and possibly other
ingredients of modern theoretical evolutionary models.
}
\keywords{
Stars: evolution --
Stars: fundamental parameters --
Stars: binaries: eclipsing --
Stars: individual: BK\,Peg, BW\,Aqr -- 
Techniques: photometric --
Techniques: spectroscopic}

\maketitle

\begin{figure*}
\epsfxsize=185mm
%\epsfbox{bkpeg_lc.ps}
\epsfbox{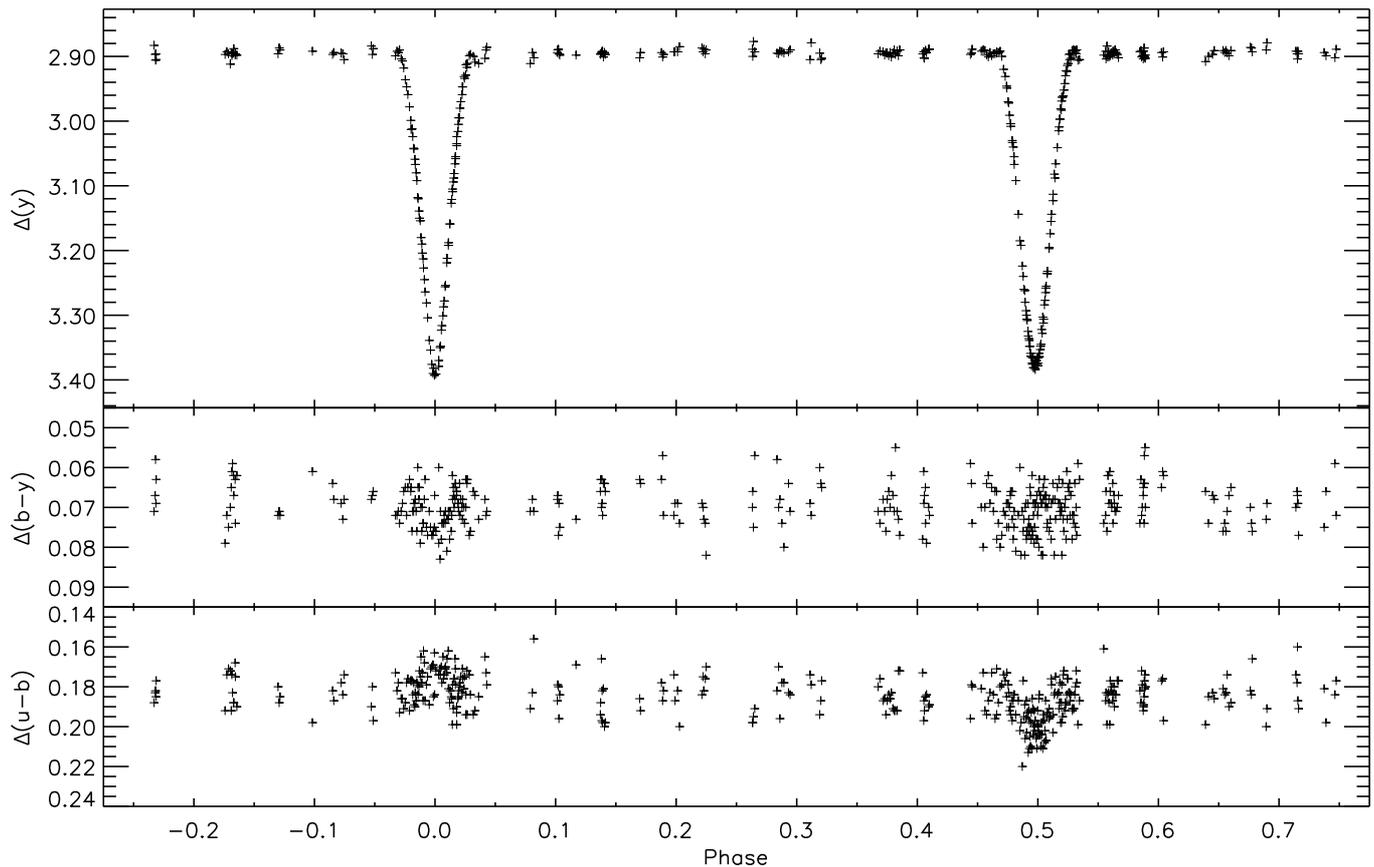}
\caption[]{\label{fig:bkpeg_lc}
$y$ light curve and $b-y$ and $u-b$ colour curves (instrumental system)
for BK\,Peg.
}
\end{figure*}

\section{Introduction}
\label{sec:intro}

Detached, double-lined eclipsing binaries (dEB) are our main source for stellar
masses and radii, today accurate to 1\% or better (Torres et al. \cite{tag09}),
and they also provide stringent
tests of various aspects of stellar evolutionary models. For this purpose,
well-established abundance information is needed, as demonstrated by
e.g. Clausen et al. (\cite{avw08}, hereafter CTB08).
One of the troublesome ingredients in theoretical models for stars 
heavier than the Sun is the amount of and treatment of convective core 
overshoot, and in this paper we focus on that aspect.

The literature on the existence and calibration of core overshoot is extensive, and
here we only draw attention to a few studies based on binary and cluster results.
From a sample of 1.5--2.5 \Msun\ dEBs and turn-off stars in IC~4651 and NGC~2680, 
Andersen et al. (\cite{anc90}) found strong evidence for convective overshoot 
in intermediate-mass stars.
Clausen (\cite{jvc91}) found indication for core overshoot for the 1.4+1.5 \Msun\
late-F type dEB \object{BW\,Aqr} and discussed \BK\ as well, and 
recently, Lacy et al. (\cite{ltc08}) found that for the 1.5 \Msun\ F7~V system 
\object{GX\,Gem}, the lowest core overshoot parameter $\alpha_{\rm ov}$ consistent with observations 
is approximately 0.18 (in units of the pressure scale height). 
The question of mass-dependence of the degree of core overshoot has -- again based on
dEB samples -- been addressed by e.g. Ribas et al. (\cite{rjg00}) and 
Claret (\cite{c07}), but they arrive at very different conclusions.
 
Most recent grids of stellar evolutionary calculations include core overshoot,
but the recipes in terms of mass and abundance dependence are somewhat different.
We refer to Sect.~\ref{sec:models} for details.

Our main motivation to undertake a study of \BK\ has been that 
a) it has evolved to the upper part of the main-sequence band and is therefore 
well-suited for core overshoot tests, but published dimensions are not 
of sufficient quality, and 
b) few similar well-studied systems are known.
Below we present absolute dimensions and abundances based on new $uvby$ light curves 
and high-resolution spectra and compare \BK\ and other similar systems with 
Yonsei-Yale and Victoria-Regina stellar evolutionary models.

In Appendix~\ref{sec:bwaqr} we present a spectroscopic abundance analysis for
BW\,Aqr as supplement to the study of this binary by Clausen (\cite{jvc91}).

\section{BK\,Peg}
\label{sec:bkpeg}

\object{BK\,Peg} (BD\,+25 5003, $m_{V}$ = 9.98, Sp. type F8, $P$ = 5\fd49),
is a well detached, double-lined eclipsing binary with 1.41 and 1.26 \Msun\ 
components in a slightly eccentric ($e$ = 0.0053) orbit.
It is unusual in the sense that the more massive, larger, and 
more luminous component is slightly cooler than the other
component.  This is due to evolution, where the more massive component has
evolved to the upper part of the main-sequence band.

The eclipsing nature of \BK\ was discovered by Hoffmeister (\cite{h31}), and 
Lause (\cite{lause35,lause37}) visually observed several times of minima.
Popper \& Dumont (\cite{dmp77}) obtained $BV$ light curves at the Palomar
and Kitt Peak observatories, which were later analysed by Popper \& Etzel
(\cite{dmppe81}). Preliminary absolute dimensions were included by Popper in his
critical review of stellar masses (Popper \cite{dmp80}), and soon after he
presented spectroscopic elements and improved absolute dimensions
(Popper \cite{dmp83}). He reached component masses accurate 
to about 1\% and radii accurate to 4\% (primary) and 7\% (secondary).
Demircan et al. (\cite{d94}) presented photoelectric $UBV$ light curves 
and an improved ephemeris, as well as absolute dimensions, which, within 
the uncertainties, agree with those by Popper; see also Popper \& Etzel (\cite{dmppe95})
for clarification on terminology.
We refer to the more massive, larger but cooler component as the 
primary ($p$) component, which, for the ephemeris we adopt (Eq.~\ref{eq:bkpeg_eph}), 
is eclipsed at phase 0.0. 

\begin{table*}
\caption[]{\label{tab:bkpeg_std}
Photometric data for \BK\ and the comparison stars.
}
\begin{minipage}{\textwidth}
\begin{center}
\begin{tabular}{lllrrrrrrrrrrrr} \hline
\hline\noalign{\smallskip}
Object&Sp. Type&Ref.   &$V$&$\sigma$&$b-y$&$\sigma$&$m_1$&$\sigma$&$c_1$ &$\sigma$&N($uvby$)&$\beta$&$\sigma$&N($\beta$)\\
\noalign{\smallskip}
\hline
\noalign{\smallskip}
%jvc090608 mean uvby corrected via meanuvby_plot.pro
 BK\,Peg  & F8$^{\mathrm{a}}$         & C10    & 9.982  & 7& 0.362 & 6& 0.143 &12& 0.456 &12&176 & 2.639   & 7 & 9\\    
          &                           & C10    &10.463  & 4& 0.369 & 7& 0.136 &11& 0.477 & 4&  6 &         &   &  \\    
\hline
%HD222249 & F5~IV      &        & 7.146  & 6& 0.309 & 4& 0.177 & 9& 0.503 & 8&119 & 2.645   & 3 & 8\\ % all
 HD222249 & F5~IV$^{\mathrm{b}}$ & C10    & 7.146  & 6& 0.309 & 4& 0.177 & 8& 0.503 & 8&118 & 2.645   & 3 & 8\\ %excl dev
          &            & O83    & 7.139  & 5& 0.317 & 3& 0.158 & 4& 0.503 & 4&  1 &         &   &  \\
          &            & O94    & 7.151  & 9& 0.317 & 3& 0.161 & 3& 0.507 & 2&  3 & 2.645   & 8 & 3\\
          &            & P69    & 7.010  &  & 0.303 &  & 0.190 &  & 0.487 &  &  2 &         &   &  \\
%         &            & P69    & 7.01   &  & 0.303 &  & 0.190 &  & 0.487 &  &  2 &         &   &  \\
%n2000                 &          7.148       0.317      0.160      0.506       4   2.645         3
%bq_prcats                                                                          2.645     8   3
%osnbeta_bprcats                                                                    2.648     5   1  not included
%
\hline
%HD222391 & G0~III     &        & 7.567  & 6& 0.362 & 6& 0.185 &11& 0.424 & 9&137 & 2.617   & 4 & 8\\
 HD222391 & G0~III$^{\mathrm{c}}$ & C10    & 7.566  & 6& 0.362 & 6& 0.185 &11& 0.424 & 8&134 & 2.617   & 4 & 8\\
          &            & O83    & 7.559  & 5& 0.371 & 3& 0.170 & 4& 0.420 & 5&  1 &         &   &  \\
          &            & O94    &        &  & 0.372 & 3& 0.167 & 4& 0.427 & 6&  1 & 2.620   &  6& 2\\
%         &            & O94    & 7.564  &  & 0.372 & 3& 0.167 & 4& 0.427 & 6&  1 & 2.620   &  6& 2\\
%n2000                 &          7.559       0.371      0.168      0.423       2   2.624         1
%bq_prcats                                                                          2.624     6   1  included
%osnbeta_bprcats                                                                    2.616     5   1  included
%
\hline
%HD223323 & F2~IV-V    &        & 7.084  & 6& 0.290 & 5& 0.138 &10& 0.414 & 9&189 & 2.636   & 7 &13\\
 HD223323 & F2~IV-V$^{\mathrm{d}}$ & C10    & 7.084  & 6& 0.290 & 5& 0.138 &10& 0.414 & 9&189 & 2.636   & 7 &13\\
          &            & J96    & 7.076  & 4& 0.302 & 1& 0.123 & 1& 0.407 & 1&  9 & 2.640   &   &  \\
          &            & J96    & 7.094  & 2& 0.300 & 0& 0.128 & 5& 0.409 & 4&  2 &         &   &  \\
          &            & J96    & 7.089  & 1& 0.299 & 0& 0.120 & 0& 0.406 & 1&  2 &         &   &  \\
          &            & O83    & 7.080  & 6& 0.296 & 5& 0.124 & 5& 0.412 & 4&  4 & 2.643   & 5 & 4\\
%n2000                 &          7.080       0.296      0.124      0.412       4   2.643         4
\hline
\end{tabular}
\begin{list}{}{}
\item[$^{\mathrm{a}}$] Popper (\cite{dmp83}), 
$^{\mathrm{b}}$ Harlan \& Taylor (\cite{ht70}), 
$^{\mathrm{c}}$ Heard (\cite{h56}), 
$^{\mathrm{d}}$ Harlan (\cite{h69}) 
\end{list}
\end{center}
\textsc{NOTE 1:}                                                                                    
References are:                                                                                     
C10 = This paper.
J96 = Jordi et al. (\cite{jordi96}). % AAS 115, 401    std star 223323, 3 values
O83  = Olsen (\cite{olsen83}).   % A&AS 54, 55
O94  = Olsen (\cite{olsen94}).   % A&AS 106, 257
P69  = Perry  (\cite{perry69}).  % AJ 74, 705

\textsc{NOTE 2:}
For \BK, the $uvby\beta$ information by C10 is the mean value outside eclipses
(first line) and during the central part of secondary eclipse (second line), 
where 98\% ($y$) of the light of the secondary component is eclipsed. 

\textsc{NOTE 3:}
N is the total number of observations used to form the mean values, and
$\sigma$ is the rms error (per observation) in mmag.
\end{minipage}
\end{table*}

\section{Photometry}
\label{sec:phot}

Below, we present the new photometric material for \BK\ and refer to Clausen 
et al.  (\cite{jvcetal01}; hereafter CHO01) for further details on observation 
and reduction procedures, and determination of times of minima.

\subsection{Light curves for BK\,Peg}
\label{sec:lc}

The differential $uvby$ light curves of \BK\ were observed at the
Str{\"o}mgren Automatic Telescope (SAT) at ESO, La Silla
with its 6-channel $uvby\beta$ photometer on 66 nights between
October 2000 and September 2003 (JD2451828--2452910). 
They contain 384 points per band with most phases covered at least twice.
The observations were done through an 18 arcsec diameter circular
diaphragm at airmasses between 1.8 and 2.2. 
\object{HD~222249}, \object{HD~222391}, and 
\object{HD~223323} -- all within a few degrees of \BK\ on the sky -- were 
used as comparison stars and were all found to be constant within a few mmag; 
see Table~\ref{tab:bkpeg_std}. 
The light curves are calculated relative to HD~223323, but all comparison 
star observations were used, shifting them first to the same light level.
HD~223323 has later been found to be a double-lined spectroscopic binary
with an estimated orbital inclination of about $63\degr$;
the orbital period is 1175 days and the orbital eccentricity is
0.6 (Griffin \cite{griffin07}). 
The average accuracy per light curve point is about 5 mmag ($ybv$) and 
8 mmag ($u$).
The light curves (Table 13) will only be available in electronic form.

As seen from Fig.~\ref{fig:bkpeg_lc}, \BK\ is well detached with 
nearly identical eclipse depths of about 0.5 mag. In the $y,b$, and $v$ 
bands, the primary eclipse at phase 0.0, which is a transit, is slightly 
deeper than the secondary eclipse (almost total occultation), which
occurs at phase 0.4976. In the $u$ band the secondary eclipse is,
however, the slightly deeper one.

\subsection{Standard photometry for \BK}
\label{sec:std}

Standard $uvby\beta$ indices for \BK\ and the three comparison stars,
observed and derived as described by CHO01, are presented in 
Table~\ref{tab:bkpeg_std}. As seen, the indices are based on many
observations and their precision is high.
For comparison, we have included published photometry from 
other sources.
In general, the agreement is good, but in some cases differences are 
larger than the quoted errors; we have used the new results 
for the analysis of \BK.

\subsection{Times of minima and ephemeris for \BK}
\label{sec:tmin}

Three times of secondary minimum, but none of primary, have been determined
from the $uvby$ light curve observations. They are listed in 
Table~\ref{tab:bkpeg_tmin} together with available measured times. 
From separate weighted least squares fit to the times of primary and
secondary minima, respectively,
we derive the linear ephemeris given in Eq.~\ref{eq:bkpeg_eph}.

\begin{equation}
\label{eq:bkpeg_eph}
\begin{tabular}{r r c r r}
{\rm Min \, I} =  & 2450706.46968 & + & $5\fd 48991046$ &$\times\; E$ \\
                  &      $\pm  38$&   &    $\pm     56$ &             \\
\end{tabular}
\end{equation}

\noindent
Within errors, the two types of minima yield identical periods, and
the new ephemeris is in good agreement with that by Demircan et al.
(\cite{d94}); see also Kreiner et al. (\cite{kreiner01}) and
Kreiner (\cite{kreiner04})\footnote{{\tt http://www.as.ap.krakow.pl/ephem}}.

\section{Spectroscopy}
\label{sec:spec}

\begin{table*}
\caption[]{\label{tab:bkpeg_tmin}
Times of primary (P) and secondary (S) minima for BK\,Peg.
%O-C values (P) and phases (S) are calculated for the
%ephemeris given in Eq. \ref{eq:bkpeg_eph}.
%References are:
%A03 = Ak et al. (\cite{ak03}).
%B01 = BBSAG Bull. 126.
%B87 = Braune \& H\"uebscher (\cite{bh87}).
%C10 = This paper.
%D94 = Demircan et al. (\cite{d94}).
%D77 = Dvorak (\cite{dworak77}).
%H07 = H\"uebscher \& Walter (\cite{hw07}).
%L35 = Lause (\cite{lause35}).
%L37 = Lause (\cite{lause37}).
%P83 = Popper \& Dumont (\cite{dmp77}), redetermined.
%Observing methods are: V = visual; PE = photoelectric; CCD = CCD.
%Observing methods are: PG = photographic; V = visual; PE = photoelectric; CCD = CCD.
}
\begin{minipage}{\textwidth}
\begin{center}
\begin{tabular}{lllrllc} \hline
\hline\noalign{\smallskip}
HJD          & rms     &  rms    &   (O-C)/phase$^{\mathrm{a}}$ & Type& Observing& Reference$^{\mathrm{c}}$  \\
$-$ 2\,400\,000& publ. &  adopt. &   days                       &     & method$^{\mathrm{b}}$&            \\
\noalign{\smallskip}
\hline
\noalign{\smallskip}
2426600.280  &  0.005  &         &  0.0072  &  P  & V        &  D77  \\%  dworak 1977  ????????????
2427418.270  &         & 0.01    &  0.0005  &  P  & V        &  L35  \\%  lause 1935, vis, rms assumed
2427429.277  &         & 0.01    &  0.0277  &  P  & V        &  -    \\
2427698.252  &         & 0.01    &$-0.0029$ &  P  & V        &  -    \\
2427709.224  &         & 0.01    &$-0.0108$ &  P  & V        &  -    \\
2427764.158  &         & 0.01    &  0.0241  &  P  & V        &  -    \\
2428373.507  &         & 0.01    &$-0.0069$ &  P  & V        &  L37  \\%  lause 1937, vis, rms assumed
2428395.473  &         & 0.01    &$-0.0006$ &  P  & V        &  -    \\
2428428.424  &         & 0.01    &  0.0110  &  P  & V        &  -    \\
2428461.348  &         & 0.01    &$-0.0045$ &  P  & V        &  -    \\
2428538.220  &         & 0.01    &  0.0088  &  P  & V        &  -    \\
2441587.7265 &   0.0010&         &$-0.0019$ &  P  & PE       &  P83  \\%  dmp 1983, pe, rms assumed
2446737.244  &         & 0.01    &$-0.0204$ &  P  & V        &  B87  \\%  BAV 1987, vis, rms assumed
2448900.28902&  0.00056&         &$-0.00012$&  P  & PE       &  D94  \\%  demircan et al. 1994, pe
2450706.4701 &   0.0003&         &  0.0004  &  P  & PE       &  A03  \\%  ibvs 5361 UBV
2451052.3325 &   0.0007&         &$-0.0015$ &  P  & PE       &  -    \\%  ibvs 5361 UBV
2454000.4181 &   0.0028&         &  0.0021  &  P  & CCD      &  H07  \\%  ibvs 5761 ccd n=37 fil = -Ir
2427684.556  &         & 0.01    &  0.5047  &  S  & V        &  L35  \\%  lause, 1935, vis, rms assumed
2427717.444  &         & 0.01    &  0.4953  &  S  & V        &  -    \\
2427739.416  &         & 0.01    &  0.4975  &  S  & V        &  -    \\
2427827.250  &         & 0.01    &  0.4968  &  S  & V        &  -    \\
2428480.560  &         & 0.01    &  0.4987  &  S  & V        &  L37  \\%  lause, 1937, vis, rms assumed
2428513.497  &         & 0.01    &  0.4982  &  S  & V        &  -    \\
2428535.438  &         & 0.01    &  0.4949  &  S  & V        &  -    \\
2428546.449  &         & 0.01    &  0.5005  &  S  & V        &  -    \\
2428557.427  &         & 0.01    &  0.5002  &  S  & V        &  -    \\
2428579.368  &         & 0.01    &  0.4968  &  S  & V        &  -    \\
2441974.7539 &   0.0003&         &  0.4976  &  S  & PE       &  P83  \\%  popper dumont redetermined
2443786.43575&  0.00025& 0.00100 &  0.49967 &  S  & PE       &  D94  \\%  demircan et al. 1994
2448886.56089&  0.00100&         &  0.49937 &  S  & PE       &  -    \\
2450319.4175 &   0.0004&         &  0.4975  &  S  & PE       &  A03  \\%  ibvs 5361 UBV
2451137.4160 &   0.0004&         &  0.4979  &  S  & PE       &  -    \\%  ibvs 5361 UBV
2451845.6124 &   0.0002&         &  0.4975  &  S  & PE       &  C10  \\%  this paper
2451867.5715 &   0.0008&         &  0.4974  &  S  & PE       &  -    \\%  this paper
2452136.580  &   0.002 &         &  0.4979  &  S  & CCD      &  D01  \\%  BBSAG 2001, ccd
2452910.6565 &   0.0002&         &  0.4978  &  S  & PE       &  C10  \\%  this paper
\hline
\end{tabular}
\begin{list}{}{}
\item[$^{\mathrm{a}}$] 
O-C values (P) and phases (S) are calculated for the
ephemeris given in Eq. \ref{eq:bkpeg_eph}.
\item[$^{\mathrm{b}}$]
Observing methods are: V = visual; PE = photoelectric; CCD = CCD.
\item[$^{\mathrm{c}}$]
References are:
A03 = Ak et al. (\cite{ak03}).
B87 = Braune \& H\"ubscher (\cite{bh87}).
C10 = This paper.
%D77 = Dvorak (\cite{dworak77}).
D77 = Dworak (\cite{dworak77}).
D94 = Demircan et al. (\cite{d94}).
%B01 = BBSAG Bull. 126.
D01 = Diethelm (\cite{bbsag01}).
H07 = H\"ubscher \& Walter (\cite{hw07}).
L35 = Lause (\cite{lause35}).
L37 = Lause (\cite{lause37}).
P83 = Popper \& Dumont (\cite{dmp77}), redetermined.
\end{list}
\end{center}
\end{minipage}
\end{table*}

\begin{table}
\caption[]{\label{tab:fies}
Log of the FIES observations of BK\,Peg.
%jvc080827
}
\begin{minipage}{\columnwidth}
\centering
\renewcommand{\footnoterule}{}  % to avoid a line before footnotes
\begin{tabular}{ccrrr}
\hline
\hline\noalign{\smallskip}
HJD$-$2\,400\,000\footnote{Refers to mid-exposure}  & phase &t$_{exp}$\footnote{Exposure time in seconds} & S/N\footnote{Signal-to-noise ratio measured around 6070 {\AA}}\\
\noalign{\smallskip}
\hline
\noalign{\smallskip}
54333.45954& 0.6647 &  300 &  60 \\ %    200047 
54333.46395& 0.6655 &  300 &  55 \\ %    200048 
54334.42028& 0.8397 &  300 &  20 \\ %    210040 
54335.52799& 0.0414 &  300 &  70 \\ %    220030 
54335.53872& 0.0434 &  600 &  85 \\ %    220032 
54335.62203& 0.0586 &  600 &  85 \\ %    220036 
54335.72881& 0.0780 &  600 & 100 \\ %    220038 
54336.41869& 0.2037 &  600 &  55 \\ %    230129 
54336.43503& 0.2067 &  600 &  60 \\ %    230131 
54336.52436& 0.2229 &  600 &  75 \\ %    230135 
54336.74168& 0.2625 &  600 &  80 \\ %    230141 
54337.44772& 0.3911 &  600 &  65 \\ %    240190 
54337.58944& 0.4169 &  600 &  75 \\ %    240196 
\hline
\end{tabular}
\end{minipage}
\end{table}

In order to perform abundance determinations and also improve the
spectroscopic elements by Popper (\cite{dmp83}), we have obtained 
13 high-resolution (R = 45000) spectra with the FIES fibre echelle 
spectrograph at Nordic Optical Telescope, La Palma during 
five consecutive nights in August 2007; 
see Table~\ref{tab:fies}. For the basic reduction of the spectra, 
we have applied the IRAF based 
FIEStool package{\footnote{see {\tt http://www.not.iac.es} for details on 
FIES and FIEStool.}. 
Subsequently,
dedicated IDL\footnote{{\tt http://www.ittvis.com/idl/index.asp}}
programs were applied to remove cosmic ray events and other defects,
and for normalisation of the individual orders. For each order, only
the central part with acceptable 
signal-to-noise ratios was kept for further ana\-ly\-sis.

\begin{table}
\caption[]{\label{tab:bkpeg_rv}
Radial velocities of BK\,Peg and residuals from the final spectroscopic orbit
presented in Table~\ref{tab:bkpeg_spel}.}
\scriptsize{
\begin{center}
\begin{tabular}{llrrrr} \hline
\hline\noalign{\smallskip}
\multicolumn{1}{c}{HJD}&\multicolumn{1}{c}{Phase}      &\multicolumn{1}{c}{$RV_p$}&\multicolumn{1}{c}{$RV_s$}&\multicolumn{1}{c}{$(O-C)_p$}&\multicolumn{1}{c}{$(O-C)_s$} \\
$-$2\,400\,000&                         &\multicolumn{1}{c}{\kms}&\multicolumn{1}{c}{\kms}&\multicolumn{1}{c}{\kms}&\multicolumn{1}{c}{\kms} \\
\hline\noalign{\smallskip}
54333.45954& 0.6647 &  60.398 &$-83.802$&  0.25   &   0.28  \\   %300 &  60 \\ %    200047
54333.46395& 0.6655 &  60.694 &$-83.936$&  0.35   &   0.38  \\   %300 &  55 \\ %    200048
54334.42028& 0.8397 &  58.578 &$-81.562$&$-0.50$  &   0.09  \\   %300 &  20 \\ %    210040
54335.52799& 0.0414 &$-28.144$&  15.916 &$-0.19$  & $-0.38$ \\   %300 &  70 \\ %    220030
54335.53872& 0.0434 &$-28.875$&  16.995 &  0.02   & $-0.33$ \\   %600 &  85 \\ %    220032
54335.62203& 0.0586 &$-36.037$&  24.823 &  0.05   & $-0.42$ \\   %600 &  85 \\ %    220036
54335.72881& 0.0780 &$-44.698$&  34.832 &  0.22   & $-0.10$ \\   %600 & 100 \\ %    220038
54336.41869& 0.2037 &$-83.237$&  77.803 &  0.15   &   0.44  \\   %600 &  55 \\ %    230129
54336.43503& 0.2067 &$-83.671$&  78.279 &  0.11   &   0.47  \\   %600 &  60 \\ %    230131
54336.52436& 0.2229 &$-85.447$&  80.223 &  0.03   &   0.45  \\   %600 &  75 \\ %    230135
54336.74168& 0.2625 &$-86.151$&  81.269 &$-0.01$  &   0.50  \\   %600 &  80 \\ %    230141
54337.44772& 0.3911 &$-56.696$&  48.134 &$-0.20$  & $-0.59$ \\   %600 &  65 \\ %    240190
54337.58944& 0.4169 &$-46.161$&  36.119 &$-0.28$  & $-0.80$ \\   %600 &  75 \\ %    240196
\noalign{\smallskip}
\hline
\end{tabular}
\end{center}
}
\end{table}

The radial velocities for \BK\
were measured from 40 useful orders of the 13 FIES spectra. 
We applied the broadening function (BF) formalism 
(Rucinski \cite{r99,r02,r04}), 
using $v \sin i = 0$ synthetic templates matching the effective 
temperature, log($g$), and metal abundance of the components of \BK. 
They were calculated with the $bssynth$ tool, 
which applies the SYNTH software (Valenti \& Piskunov 
\cite{vp96}) and modified ATLAS9 models (Heiter et al. \cite{heiter02}). 
Since the components have nearly identical temperatures, the two templates
are very similar and lead to practically identical results.
As described by e.g. Kaluzny et al. (\cite{k06}), the projected rotational
velocities $v \sin i$  of the components and (monochromatic) 
light/luminosity ratios between them can also be obtained from analyses 
of the BFs. 

For each spectrum, BFs were calculated for each of the selected orders,
and a mean BF was then calculated together with weights for
each order based on the root mean square deviation of the
individual BFs from the mean BF. 
The final BF is the weighted average for the selected orders. 
The radial velocites, the $v \sin i$'s, and the light 
ratio were derived by fitting a rotational profile for both stellar components,
convolved with a Doppler profile corresponding to the instrumental
resolution, to the final BF for each observed spectrum. 

The radial velocities are listed in Table~\ref{tab:bkpeg_rv}.
The final values of $v\sin i$ and light 
ratio were calculated as the mean values for the 13 spectra,
with errors  estimated from the deviations from spectrum to spectrum. 
For the primary and secondary components of \BK, we obtain mean rotational 
velocities of $16.6 \pm 0.2$ and $13.4 \pm 0.2$ \kms, respectively.
For the light ratio we find $L_s/L_p = 0.57 \pm 0.02$.

In addition, we have determined the light ratio between the components
by directly comparing the FIES spectra and synthetic binary spectra, 
calculated for a range of luminosity ratios between the components. 
Adopting the temperatures, surface gravities,
rotational velocities, and metallicities listed in Table~\ref{tab:bkpeg_absdim},
and using several spectral orders covering 5300--5800 {\AA}, we obtain the best 
line fits for a light ratio of $0.57 \pm 0.04$. 
As expected, since the components have nearly identical temperatures, we find
no significant wavelength dependence of the spectroscopic light ratio, 
even if a broader wavelength region is used. 

\begin{table*}
\caption[]{\label{tab:teff}
%Effective temperatures (K) for the `average' component of \BK.
Effective temperatures (K) for the combined light of \BK.
}
%\scriptsize{
\begin{center}
\begin{tabular}{ccccccccccc} \hline
\hline\noalign{\smallskip}
 $A_V$& [Fe/H] & $(b-y)_0$ &$c_0$&$(V-J)_0$&$(V-H)_0$&$(V-K_s)_0$& A96 & H07 &RM05 & M06  \\
\noalign{\smallskip}
\hline
\noalign{\smallskip}
%0.188&$-0.10$ &0.405&0.320&1.208&1.458&1.564 & 6280 & 6270 &6270/6255/6130/6265 &  6290 \\
 0.188&$-0.12$ &0.405&0.320&1.208&1.458&1.564 & 6280 & 6270 &6270/6255/6130/6265 &  6290 \\
%         .064    .20                                    150    120        145  110  110
\noalign{\smallskip}            
\hline
\end{tabular}            
\end{center}    
\textsc{Note 1:}
$A_V$ is the adopted visual interstellar absorption.
The $V_0$ magnitude and the $(b-y)_0$ and $c_0$ indices are based on the 
out-of-eclipse $uvby$ standard indices from Table~\ref{tab:bkpeg_std}.
The 2MASS observations ($J,H,K_s$) were obtained at phase 0.879.

\textsc{Note 2:}
References are: 
A96 = Alonso et al. (\cite{alonso96}). 
H07 = Holmberg et al. (\cite{holmberg07}).
RM05 = Ram\'\i rez \& Mel\'endez (\cite{rm05}). 
M06 = Masana et al. (\cite{masana06}).

\textsc{Note 3:}
The results from A96 are based on their $uvby$ calibration,
those from RM05 on their $uvby$, $(V-J)$, $(V-H)$, and $(V-K_s)$ calibrations (in that order); 
the calibration by M06 is for $(V-K_s)$. 
%}
\end{table*}

\section{Photometric elements}
\label{sec:phel}

Since \BK\ is well-detached, the photometric elements have been determined
from {\sc jktebop}\footnote{{\tt http://www.astro.keele.ac.uk/$\sim$jkt/}} 
analyses (Southworth et al. \cite{sms04a},\cite{sms04b}) of the  $uvby$ light curves.
The underlying simple Nelson-Davis-Etzel binary model (Nelson \& Davis \cite{nd72},
Etzel \cite{e81}, Popper \& Etzel  \cite{dmppe81}, Martynov \cite{m73}) 
represents the deformed stars as biaxial ellipsoids and applies a simple 
bolometric reflection model. We refer to CTB08 
for details on 
the general approach applied.
In tables and text, we use the following symbols:
$i$ orbital inclination;
$e$ eccentricity of orbit;
$\omega$ longitude of periastron;
$r$ relative radius;
$k = r_s/r_p$;
$u$ linear limb darkening coefficient;
$y$ gravity darkening coefficient;
$J$ central surface brightness;
$L$ luminosity;
$T_{\rm eff}$ effective temperature.

The mass ratio between the components was kept at the spectroscopic value,
see Sect.~\ref{sec:spel}.
The simple built-in bolometric reflection model was used, 
linear limb darkening coefficients by Van Hamme (\cite{vh93}) and  
Claret (\cite{c00}) were applied, and
gravity darkening coefficients corresponding to radiative atmospheres
were adopted. 
Identical coefficients were used for the two components, since their effective
temperatures and surface gravities are sufficiently identical. 
Effective temperatures determined from
the standard $uvby$ and $JHK_s$ indices outside eclipses are listed
in Table~\ref{tab:teff}. As seen, the results from the different
calibrations agree well; we have adopted the temperature based on the
Holmberg et al. (\cite{holmberg07}) calibration.

Solutions for \BK, based on Van Hamme limb darkening coefficients,
are presented in Table~\ref{tab:bkpeg_jktebop_vh},
and $O\!-\!C$ residuals of the $y$ observations from the theoretical
light curve are shown in Fig.~\ref{fig:bkpeg_res_y}.
As seen, the results from the four bands agree well.
Changing to Claret (\cite{c00}) limb darkening coefficients,
which are 0.07--0.09 higher, increases the radius of the primary component
by only 0.4\%, whereas that of the secondary component is increased
by 1.5\%. This is linked to a 1\% larger $k$ and a 29\% smaller 
$e\mathrm{sin}(\omega)$, reducing $e$ by 10\%.
Limb darkening coefficients determined from the light curves reproduce those by Van 
Hamme better than those by Claret, but have uncertainties of about $\pm 0.12$.
Including non-linear limb darkening (logarithmic or
square-root law) has no significant effect on the photometric elements.

The adopted photometric elements listed in Table~\ref{tab:bkpeg_phel}
are the weighted mean values of the {\sc jktebop} solutions adopting the linear
limb darkening coefficients by Van Hamme. 
Realistic errors, based on 10\,000 Monte Carlo simulations in each band 
and on comparison between the $uvby$ solutions, have been assigned.
The Monte Carlo simulations include random variations within $\pm 0.07$
of the linear limb darkening coefficients.
As seen, $r_p$ becomes more accurate than $r_s$. This is because it correlates
less with $k$, probably related to the secondary eclipse being nearly total
for the adopted elements.
It should be noted that the $ybv$ luminosity ratios from the
light curve solutions agree very well with the spectroscopic light ratio 
(Sect.~\ref{sec:spec}). 

\begin{table}
\caption[]{\label{tab:bkpeg_jktebop_vh}
Photometric solutions for BK\,Peg from the {\sc jktebop} code.
%jvc 080825        new eph
%5.48991046        Orbital period of eclipsing binary system (days)
%50706.46968       Reference time of primary minimum (HJD)
%rej. sigma 3.0 included
}
\begin{center}
\begin{tabular}{lrrrr} \hline
\hline\noalign{\smallskip}
                     &     $y$    &       $b$  &       $v$  &   $u$\\                   
\noalign{\smallskip}
\hline
\noalign{\smallskip}
$i$ \, (\degr)       &  88.02     &   87.91    &   87.99    &   87.90\vspace{-0.8mm}\\   
                     & $\pm 5$    &  $\pm 5$   &  $\pm 4$   &  $\pm 7$\\                 
%c                      87.85         87.72        87.85        87.72
%sqrt                   87.99
%log                    87.99

$e\cos \omega$       &$-0.00364$  & $-0.00376$ & $-0.00363$ & $-0.00374$\vspace{-0.8mm}\\ 
                     &$\pm    6$  & $\pm    7$ & $\pm    6$ & $\pm   10$\\             
%c                     -0.00364      -0.00376     -0.00363     -0.00375
%sqrt                  -0.00364
%log                   -0.00364

$e\sin \omega$       &$ 0.00283$  & $ 0.00563$ & $ 0.00438$ & $ 0.00059$\vspace{-0.8mm}\\  
                     & $\pm 181$  &  $\pm 207$ & $\pm  183$ & $\pm  312$\\                
%c                      0.00133       0.00504      0.00340     -0.00050
%sqrt                   0.00264
%log                    0.00264

$e$                  &  0.0046    &   0.0068   &   0.0057   &  0.0038\\                 
%c                      0.0038        0.0063       0.0050      0.0038          
%sqrt                   0.0045
%log                    0.0045
                                                         
$\omega$ \, (\degr)  &  142.1     &   123.8    &   129.6    & 170.0  \\                 
%c                      160.0         126.7        136.9      187.7       
%sqrt                   144.0
%log                    144.0          
                            
$r_p$                &  0.1092    &   0.1096   &   0.1092   &  0.1100\\                 
%c                      0.1096        0.1098       0.1095      0.1104
%sqrt                   0.1094
%log                    0.1093

$r_s$                &  0.0806    &   0.0819   &   0.0807   &  0.0813\\                 
%c                      0.0817        0.0834       0.0817      0.0826
%sqrt                   0.0807
%log                    0.0807          
                            
$k$                  &  0.7379    &   0.7474   &   0.7394   &   0.7391\vspace{-0.8mm}\\  
                     &  $\pm50$   &   $\pm68$  &   $\pm46$  &   $\pm97$\\                
%c                      0.7461        0.7595       0.7456       0.7482
%sqrt                   0.7379
%log                    0.7380

$r_p + r_s$          &  0.1898    &   0.1915   &   0.1900   &  0.1913\vspace{-0.8mm}\\  
                     &  $\pm 5$   &   $\pm 5$  &   $\pm 4$  &  $\pm 7$ \\               
%$r_p + r_s$          &  0.18983   &   0.19147  &   0.19000  &  0.19130\vspace{-0.8mm}\\  
%                     &  $\pm 46$  &   $\pm 51$ &   $\pm 44$ &  $\pm 74$ \\               
%c                      0.19130       0.19312      0.19120     0.19292            
%sqrt                   0.19005
%log                    0.19004          
                            
$u_p = u_s$          &  0.55      &   0.64     &   0.72     &  0.70\\                   
%c                      0.64          0.73         0.79        0.79

$y_p = y_s$          &  1.09      &   1.24     &   1.42     &  1.68\\

$J_s/J_p$            &  1.0444    &    1.0447  &   1.0569   &  1.1016\vspace{-0.8mm}\\
                     &  $\pm31$   &    $\pm37$ &   $\pm34$  &  $\pm60$\\
%c                      1.0456         1.0437      1.0567      1.1007
%sqrt                   1.0446
%log                    1.0444

$L_s/L_p$            &  0.5670    &    0.5817  &   0.5756   &  0.5990 \\
%c                      0.5804         0.6002      0.5852      0.6133
%sqrt                   0.5670
%log                    0.5671

$\sigma$ \, (mmag.)  &  4.8       &    5.2     &   4.7      &  7.9   \\
%c                      4.8            5.2         4.7         7.9
%sqrt                   4.8
%log                    4.8

% phase shift          -0.000016      -0.000002   -0.000022   -0.000017                
%                                            
\noalign{\smallskip}            
\hline
\end{tabular}            
\end{center}            
\textsc{Note 1:}
Linear limb darkening coefficients by Van Hamme (\cite{vh93}) were adopted,
a mass ratio of 0.89 was assumed, and phase shift and  mag. normalization
%were included as free patameters
were included as free parameters

\textsc{Note 2:}
The errors quoted for the free parameters are the $formal$ errors determined
from the iterative least squares solution procedure
\end{table}                       

\begin{table}            
\caption[]{\label{tab:bkpeg_phel}
Adopted photometric elements for BK\,Peg.
%jvc 080827  
}
%\begin{flushleft}             
\begin{center}             
\begin{tabular}{ll}             
\noalign{\smallskip}             
\hline             
\noalign{\smallskip}             
%       ebop_vh..mean of uvby                         claret limb     P&E 1981      Dem
$i$              & $87{\fdg}96 \pm 0{\fdg}14$ \\      %  87.79        87.5  .2      88.05  0.37
$e$              & $0.0053 \pm 0.0013$ \\             %  0.0048
$\omega$         & $138{\fdg}7 \pm 19{\fdg}0$ \\      %  149.4
%ecosw            -0.003685                             -0.003687 
%esinw             0.003620                              0.002576
$r_p$            & $0.1094 \pm 0.0004$ \\             %  0.1098       0.108  .004   0.110  .006
$r_s$            & $0.0811 \pm 0.0009$ \\             %  0.0823       0.086  .006   0.085  .006
$r_p + r_s$      & $0.1906 \pm 0.0012$ \\             %  0.1920
$k$              & $0.741\pm 0.008$ \\                %  0.749
\noalign{\smallskip}             
\end{tabular}             
\begin{tabular}{lrrrr}             
\noalign{\smallskip}             
                 & $y$    & $b$    & $v$   & $u$  \\           
\noalign{\smallskip}             
$J_s/J_p$        & 1.044  & 1.048  & 1.057 & 1.095 \\  %  for mean elements
                 &$\pm 4$ &$\pm 4$ &$\pm 4$&$\pm 6$\\   % from jktebop
$L_s/L_p$        & 0.572  & 0.574  & 0.578 & 0.599\vspace{-0.8mm} \\  % for mean elements  %P&E 0.70 DEM 0.61
                 & $\pm 11$&$\pm 14$&$\pm 10$ &$\pm 15$\\   % from jktebop
%$J_s/J_p$        & 1.0443 & 1.0481 & 1.0566& 1.0954\\  %  for mean elements
%                 & $\pm38$&$\pm38$&$\pm36$ &$\pm64$\\   % from jktebop
%$L_s/L_p$        & 0.5717 & 0.5735 & 0.5779& 0.5988\vspace{-0.8mm} \\  % for mean elements  %P&E 0.70 DEM 0.61
%                 & $\pm 108$&$\pm 142$&$\pm  96$ &$\pm 154$\\   % from jktebop
\noalign{\smallskip}             
\hline             
\end{tabular}             
%\end{flushleft}            
\end{center}            
\textsc{Note:} The individual flux and luminosity ratios are based
on the mean stellar and orbital parameters
\end{table}

For comparison, Popper \& Etzel (\cite{dmppe81}) obtained 
$r_p = 0.108 \pm 0.04$ and $r_s = 0.086 \pm 0.006$, 
assuming $e\mathrm{sin}(\omega) = 0$ and adopting $k = 0.80 \pm 0.05$
in order to reproduce a mean ratio of $0.69 \pm 0.03$ between selected 
secondary and primary lines, measured on photographic spectra. 
This ratio is, however, much higher than the light ratio we derive from 
the FIES spectra, leading to a higher $k$. Popper \& Etzel also
obtain a somewhat larger value for $r_p + r_s$.
The relative radii presented by Demircan et al. (\cite{d94}) are close to
those by Popper \& Etzel.

In conclusion, the new photometric elements derived from analyses of the 
$uvby$ light curves are significantly more accurate than previous 
determinations. 
We find that the secondary eclipse is almost total, with 98\% of the $y$ light 
of the secondary component eclipsed, whereas about
57\% of the $y$ light from the primary component is eclipsed at phase 0.0.
 
\begin{figure}
\epsfxsize=85mm
%\epsfbox{bkpeg_res_y.ps}
\epsfbox{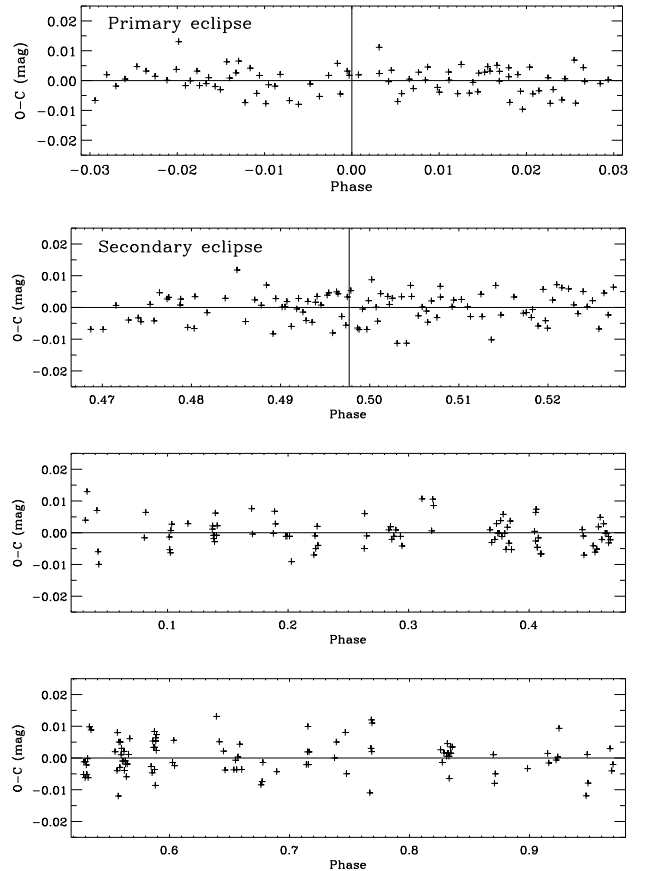}
\caption[]{\label{fig:bkpeg_res_y}
($O\!-\!C$) residuals of the \BK\ $y$-band observations from the theoretical
light curve computed for the photometric elements given in
Table~\ref{tab:bkpeg_jktebop_vh}.
}
\end{figure}

\section{Spectroscopic elements}
\label{sec:spel}

\begin{table}   
\caption[]{\label{tab:bkpeg_spel}
Spectroscopic orbital solutions for \BK.
%$T$ is the time of central primary eclipse.}
% and $T_p$ the time of periastron passage.}
%jvc 091216
}
\begin{center}    
\begin{tabular}{lrr} \hline   
\hline\noalign{\smallskip}    
%\noalign{\smallskip}    
Parameter:           & \multicolumn{1}{c}{Popper}  & \multicolumn{1}{c}{FIES} \\ 
                     &  (re-analysis)              & (adopted) \\
\noalign{\smallskip}
\hline
\noalign{\smallskip}    
Adjusted quantities:            &                  &                   \\      %dmp83
$K_p$~(\kms)                    &$ 79.14 \pm 0.33$ &  $78.77 \pm 0.11$  \\     % 79.1 0.3
$K_s$~(\kms)                    &$ 88.86 \pm 0.56$ &  $88.59 \pm 0.21$  \\     % 88.9 0.5
$\gamma_p$~(\kms)               &$ -8.51 \pm 0.30$ &  $-7.39 \pm 0.08$  \\     % -8.8 0.3
$\gamma_s$~(\kms)               &$ -8.35 \pm 0.51$ &  $-7.20 \pm 0.15$  \\     % -8.7 0.5
\noalign{\smallskip}  
Adopted quantities:                    &                  & \\
$P$~(days)                             &  5.48991046      &  5.48991046 \\ 
%$T$~(HJD$-$2\,400\,000)$^{\mathrm{a}}$ &  50706.46968     & 52719.59767 \\  
$T$~(HJD$-$2\,400\,000)$^{\mathrm{a}}$ &  50706.46968     & 50706.46968 \\  
%$T_p$~(HJD$-$2\,400\,000)       &  48775.56910     & 50365.03230 \\  
$e$                                    &  0.0053          & 0.0053      \\ 
$\omega$ \, (\degr)                    &  138.7           & 138.7       \\ 
\noalign{\smallskip}  
Derived quantities:               &                    &          \\
$M_p \sin^3i~\mathrm{(M_{\sun})}$ &$1.427  \pm 0.019 $ &$1.411 \pm 0.007$ \\
$M_s \sin^3i~\mathrm{(M_{\sun}})$ &$1.271  \pm 0.013 $ &$1.255 \pm 0.005$ \\
$q = M_s/M_p$                     & $0.891 \pm 0.007$  & $0.889 \pm 0.002$ \\
%$a_p \sin i$~($10^6$~km)          &$11.578 \pm 0.021 $ &$11.529 \\
%$a_s \sin i$~($10^6$~km)          &$11.667 \pm 0.021 $ &$11.541 \\
$a \sin i~\mathrm{(R_{\sun})}$    &$18.231 \pm 0.071 $ &$18.161 \pm 0.026$\\
\noalign{\smallskip}  
Other quantities                         &      &      \\
pertaining to the fit:                   &      &      \\
$N_{obs} (p/s)$                          &24/23 &13/13 \\
Time span (days)                         &  1374&    4 \\
$\sigma_p$$^{\mathrm{b}}$ \,(\kms)       &  1.36& 0.26 \\      %1.3
$\sigma_s$$^{\mathrm{b}}$ \,(\kms)       &  2.30& 0.50 \\      %2.3
\noalign{\smallskip}  
\hline
\end{tabular}            
\begin{list}{}{}
%\item[$^{\mathrm{a}}$] Time of central primry eclipse
\item[$^{\mathrm{a}}$] Time of central primary eclipse
\item[$^{\mathrm{b}}$] Standard deviation of a single radial velocity
\end{list}{}{}
\end{center}            
\end{table}

Spectroscopic orbits have been derived from a re-analysis of the
radial velocities by Popper (\cite{dmp83}) and an analysis of the
new radial velocities listed in Table~\ref{tab:bkpeg_rv}.
We have used the method of Lehman-Filh\'es implemented in the 
{\sc sbop}\footnote{Spectroscopic Binary Orbit Program, \\ 
{\tt http://mintaka.sdsu.edu/faculty/etzel/}}
program (Etzel \cite{sbop}), which is a modified and expanded version of
an earlier code by Wolfe, Horak \& Storer (\cite{wolfe67}).
The orbital period $P$ was fixed at the ephemeris value 
(Eq.~\ref{eq:bkpeg_eph}),
and the eccentricity $e$ and longitude of periastron $\omega$ to
the results from the photometric analysis (Table~\ref{tab:bkpeg_phel}).
The radial velocities of the components were analysed independently
(SB1 solutions).

The spectroscopic elements are presented in Table~\ref{tab:bkpeg_spel}.
The semiamplitudes ($K_p$,$K_s$), and their uncertainties, obtained from Popper's 
velocities are identical to his results, even though he assumed 
the orbit to be circular and used an older ephemeris. 

As seen, significantly more accurate semiamplitudes are derived from the 
FIES velocities. They are slightly smaller than those from Popper's 
velocities, but within errors the results agree. 
Including $e$ and/or $\omega$ as free parameters formally improves
the solution but does not alter the semiamplitudes. The number of
velocities is, however, too small for reliable spectroscopic determination
of $e$ and $\omega$. 
The double-lined (SB2) solutions agree perfectly with the single-lined 
solutions listed in Table~\ref{tab:bkpeg_spel}. 
Also, spectroscopic elements determined as part of the spectral
disentangling (Sect.~\ref{sec:abund}) are identical.
We notice that the new system velocities ($\gamma_p$,$\gamma_s$)
differ by about 1 \kms\ from 
Popper's results. This is probably due to radial velocity zero point 
differences.
Our velocities are tied to the ThAr exposures taken before and/or after
each target exposure. Standard star observations normally agree to
within 0.1--0.2 \kms.

\begin{figure}
\epsfxsize=095mm
%\epsfbox{bkpeg_sporb.ps}
\epsfbox{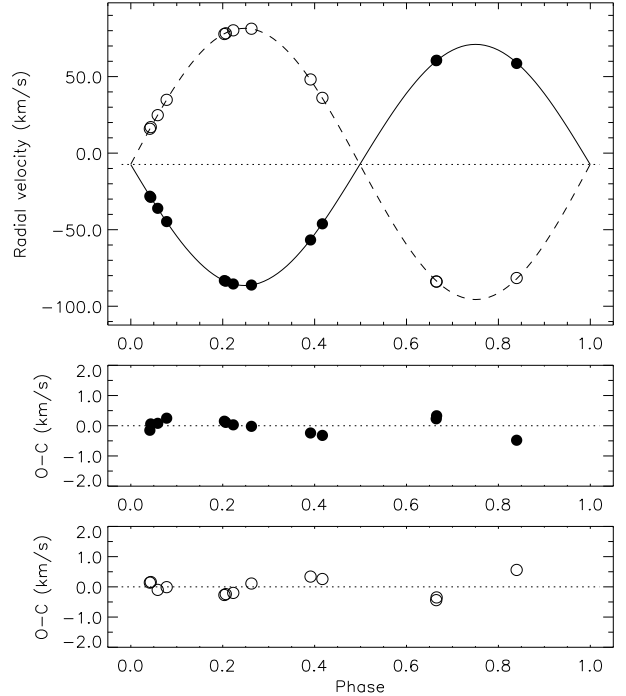}
\caption[]{\label{fig:bkpeg_sporb}
Spectroscopic orbital solution for BK\,Peg (solid line: primary;
dashed line: secondary) and radial velocities (filled circles:
primary; open circles: secondary).
The dotted line (upper panel) represents the center-of-mass
velocity of the system.
Phase 0.0 corresponds to central primary eclipse.
}
\end{figure}

\section{Chemical abundances}
\label{sec:abund}

For abundance analyses, we have disentangled the FIES spectra of \BK\
in order to extract the individual component spectra. 
We have applied the disentangling method introduced by Simon \& Sturm
(\cite{ss94}) and a revised version of the corresponding original code
developed by E.~Sturm. It assumes a constant light level, but since
\BK\ is constant to within 0.5\% outside of eclipses this is
of no concern.
Twenty-two orders, covering 5160--6450 {\AA} (with a few gaps) 
were selected and disentangled individually.
The orbital elements were fixed at the adopted values (Table~\ref{tab:bkpeg_spel})
and very slightly wavelength dependent light ratios matching the results
from the light curve analyses (Table~\ref{tab:bkpeg_phel}) were adopted.
Around 6070 {\AA}, the signal-to-noise ratios of the resulting
component spectra are 160 (primary) and 80 (secondary).
A 40~\AA\ region, centred at 6070~\AA, is shown in 
Fig.~\ref{fig:bkpeg_disentangled}

\begin{figure*}
\epsfxsize=180mm
%\epsfbox{bkpeg-entangled.ps}
\epsfbox{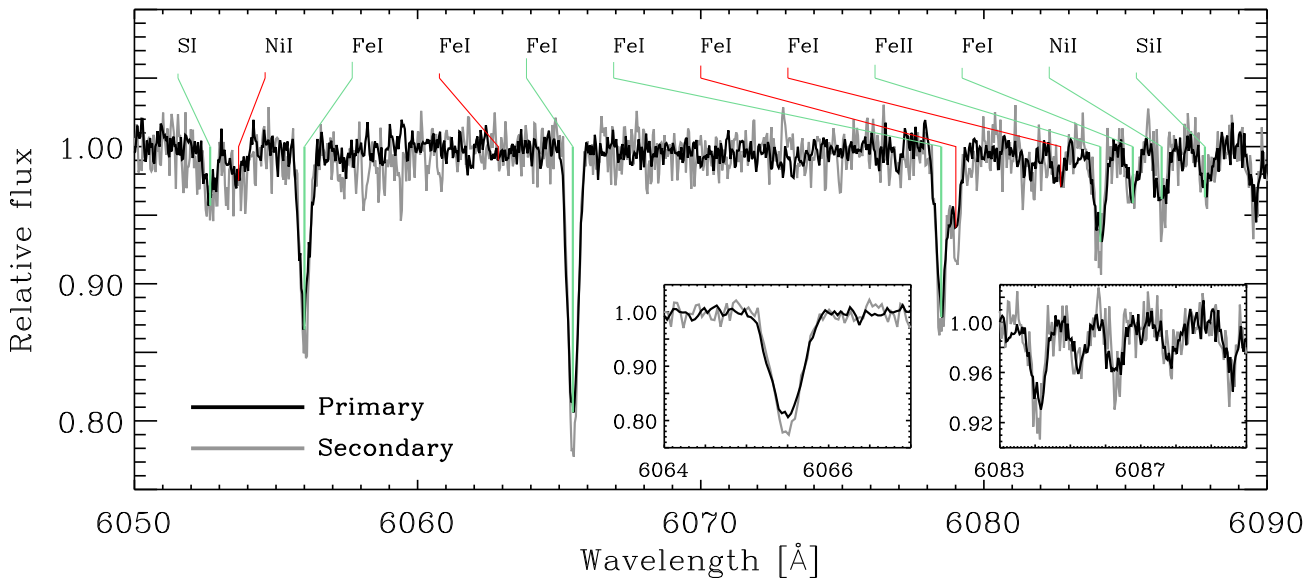}
\caption[]{\label{fig:bkpeg_disentangled}
A 40~\AA\ region centred at 6070~\AA\ of the disentangled spectra of the
components of \BK. Lines identified by a red line were not used for
the abundance analysis.
}
\end{figure*}

The basic approach followed in the abundance analyses is described
by CTB08.
We used the versatile VWA tool, which applies the SYNTH software
(Valenti \& Piskunov \cite{vp96}) to compute synthetic spectra. 
We refer to Bruntt et al.\ (\cite{bruntt04,bruntt08}) and 
Bruntt (\cite{bruntt09}) for a detailed description of VWA.
Atmosphere models were interpolated from the recent grid of MARCS
model atmospheres (Gustafsson et al. \cite{marcs08}), which adopt
the solar composition by Grevesse et al. (\cite{gas07}).
Atomic line data are from the Vienna Atomic Line Database (VALD;
Kupka et al. \cite{kupka99}), but in order to derive abundances relative
to the Sun, log($gf$) values have been adjusted
in such a way that each measured line in the Wallace et al. (\cite{whl98})
Solar atlas reproduces the atmospheric abundances by
Grevesse et al. (\cite{gas07}). 

The abundance results derived from all useful lines with equivalent
widths between 10 and 100 m{\AA} are presented in Table~\ref{tab:bkpeg_abund}.
The equivalent widths measured in the disentangled spectra are listed
in Tables~14 (primary) and 15 (secondary), which will only be available
in electronic form.
The surface gravities and observed rotational velocities listed in
Table~\ref{tab:bkpeg_absdim} were adopted, whereas the effective temperatures
and microturbulence velocities were tuned until Fe \ione\ abundances were
independent of line equivalent widths and excitation potentials.
The resulting temperatures are $6365 \pm 50$~K (primary) and
$6385 \pm 50$~K (secondary).
From a study of 10 F-K type stars with interferometrically and 
spectroscopically determined effective temperatures, 
Bruntt et al. (\cite{bruntt10})} 
find a systematic offset of 40~K, which should be subtracted.
The corrected spectroscopic temperatures are still
slightly higher than derived from the $uvby$ indices 
(Table~\ref{tab:bkpeg_absdim}) but agree within errors.

Microturbulence velocities of $1.55 \pm 0.25$ \kms\ (primary) and 
$1.22 \pm 0.25$ \kms\ (secondary) were obtained.
The calibration by Edvardsson et al. (\cite{be93}) predicts higher values of
$2.20 \pm 0.31$ \kms\ (primary) and $1.95 \pm 0.31$ \kms\ (secondary);
the difference in microturbulence will be discussed by Bruntt et al.
(\cite{bruntt10}).

\begin{table}
\caption[]{\label{tab:bkpeg_abund}
Abundances ($[\mathrm{El./H}]$) for the primary and secondary
components of \BK.
%Teff 6365, 6385 K. (man03)
%First set: man03 after q'ing .tst files.
%Second set: man03 same lines in both components .sam files
}
%$\alpha$-elements are marked by an asterix.
%jvc090701    10-100 mA, 0-10 eV, man03  i.e. Teff 6365, 6385 K   after line q  .tst files   
%             !!!! still preliminary 
\begin{minipage}{\columnwidth}
\begin{center}
\begin{tabular}{lrlrrlr} \hline
\hline\noalign{\smallskip}
             &  \multicolumn{3}{c}{Primary} & \multicolumn{3}{c}{Secondary} \\
Ion          &[El./H]&  rms& N$^{\mathrm{a}}$     &[El./H]& rms & N$^{\mathrm{a}}$            \\
\noalign{\smallskip}
\hline
Si\ione\     &$-0.19$& 0.05& 12    &$-0.12$& 0.09& 10   \\  %alpha   man03 .tst files
Ca\ione\     &$-0.05$& 0.05&  4    &  0.00 & 0.07&  6   \\  %alpha 
Sc\itwo\     &$-0.14$& 0.09&  3    &$-0.18$& 0.09&  3   \\  
Ti\ione\     &       &     &       &$-0.17$& 0.06&  3   \\  %alpha   
Ti\itwo\     &$-0.07$& 0.09&  3    &  0.01 & 0.18&  3   \\  %alpha   
Cr\ione\     &$-0.20$& 0.03&  3    &$-0.19$& 0.03&  3 \\  %          
Cr\itwo\     &$-0.23$& 0.05&  3    &$-0.04$& 0.03&  4 \\  %          
Fe\ione\     &$-0.12$& 0.07& 78    &$-0.12$& 0.09& 70      \\
Fe\itwo\     &$-0.15$& 0.09& 10    &$-0.11$& 0.11& 12 \\
Ni\ione\     &$-0.23$& 0.08&  6    &$-0.24$& 0.13& 11   \\    %      
%\hline
%Si\ione\     &$-0.18$& 0.04&  9    &$-0.12$& 0.09&  9   \\  %alpha  man03 .sam files
%Ca\ione\     &$-0.05$& 0.05&  4    &$-0.01$& 0.05&  4   \\  %alpha 
%Sc\itwo\     &$-0.14$& 0.09&  3    &$-0.18$& 0.09&  3   \\  
%Cr\ione\     &$-0.20$& 0.03&  3    &$-0.19$& 0.03&  3 \\  %          
%Cr\itwo\     &$-0.23$& 0.05&  3    &$-0.03$& 0.03&  3 \\  %          
%Fe\ione\     &$-0.13$& 0.07& 56    &$-0.11$& 0.09& 56      \\
%Fe\itwo\     &$-0.15$& 0.09& 10    &$-0.13$& 0.11& 10 \\
%Ni\ione\     &$-0.22$& 0.10&  4    &$-0.19$& 0.17&  4   \\    %      
\noalign{\smallskip}
\hline
\end{tabular}            
\begin{list}{}{}
\item[$^{\mathrm{a}}$] Number of lines used per ion
\end{list}
\end{center}
\end{minipage}
\end{table}

As seen, a robust \feh\ is obtained, with nearly identical
results from Fe\,\ione\ and Fe\,\itwo\ lines of both components.
The mean value from all measured Fe lines is \feh\,$=-0.12\pm0.01$
(rms of mean).
Changing the  model temperatures by $\pm 50$~K modifies \feh\
from the Fe\,\ione\ lines by about $\pm0.05$ dex, whereas almost no
effect is seen for Fe\,\itwo\ lines.
If 0.25 \kms\ higher microturbulence velocities are adopted,
\feh\ decreases by about 0.04 dex for both neutral and ionized lines.
Taking these contributions to the uncertainties into account,
we adopt \feh\,$=-0.12\pm0.07$ for \BK.
In general, we find similar relative abundances 
for the other ions listed in Table~\ref{tab:bkpeg_abund}, including the
$\alpha$-elements Si, Ca, and Ti.

As an addition to the spectroscopic abundance analysis, we have
also calculated metal abundances from the de-reddened $uvby$ indices for the
individual components (Table~\ref{tab:bkpeg_absdim}) and
the calibration by Holmberg et al. (\cite{holmberg07}).
The results are: \feh\,$= -0.07\pm0.18$ (primary) and
\feh\,$=-0.14\pm0.20$ (secondary). Within errors they agree with
those from the spectroscopic analysis;
the quoted \feh\ errors include the uncertainties of the
photometric indices and the published spread of the calibration.

\begin{table}   
\caption[]{\label{tab:bkpeg_absdim}
Astrophysical data for BK\,Peg.
% jvc 080828, 091216 (paramfl_new)
}
\begin{minipage}{\columnwidth}
\begin{center}    
\begin{tabular}{lrr} \hline    
\noalign{\smallskip}    
\hline    
\noalign{\smallskip}    
                     &    Primary       &    Secondary      \\ 
\noalign{\smallskip}    
\hline    
\noalign{\smallskip}    
Absolute dimensions:          &                   &                 
 \\ 
$M/M_{\sun}$                  &$1.414 \pm 0.007$  &$1.257 \pm 0.005$
\\ 
$R/R_{\sun}$                  &$1.988 \pm 0.008$  &$1.474 \pm 0.017$ 
\\ 
$\log g$ (cgs)                & $3.992 \pm 0.004$ & $4.201 \pm 0.010$
\\
%$v \sin i$ (\kms)             & $14.1\pm0.1$        & $12.1\pm0.2$   %from w1,2 in srf/vrot.jvc
$v \sin i$$^{\mathrm{a}}$ (\kms)  & $16.6 \pm 0.2$    & $13.4 \pm 0.2$ \\  %rsrf080930
$v_{sync}$$^{\mathrm{b}}$ (\kms)  & $18.3 \pm 0.1$    & $13.6 \pm 0.2$ \\ 
$v_{psync}$$^{\mathrm{c}}$ (\kms) & $18.3 \pm 0.1$    & $13.6 \pm 0.2$ \\ 
$v_{peri}$$^{\mathrm{d}}$ (\kms)  & $18.5 \pm 0.1$    & $13.7 \pm 0.2$ \\ 
 & & \\ 
Photometric data:             &                   &                 
\\ 
$V$$^{\mathrm{e}}$     &     $10.473 \pm 0.009$  &        $11.080 \pm 0.014$\\  %jvc080828 
$(b-y)$$^{\mathrm{e}}$ &     $ 0.363 \pm 0.007$  &        $ 0.360 \pm 0.008$\\
$m_1$$^{\mathrm{e}}$   &     $ 0.144 \pm 0.013$  &        $ 0.139 \pm 0.016$\\
$c_1$$^{\mathrm{e}}$   &     $ 0.466 \pm 0.014$  &        $ 0.436 \pm 0.017$\\
$E(b-y)$    & \multicolumn{2}{c}   {$0.044 \pm 0.015$} \\
%          &              &               \\
$T_{\mbox{\scriptsize eff}}\,$      &  $6265 \pm  85$    &   $6320 \pm  90$ \\
$M_{\mbox{\scriptsize bol}}\,$      &  $2.90  \pm 0.06$  &   $3.51  \pm 0.07$ \\
$\log L/L_{\sun}$ & $0.74 \pm 0.02$ &    $0.49 \pm 0.03$ \\
$B.C.$            & $-0.01$         &    $-0.01$ \\
$M_V$ &             $ 2.91 \pm 0.06$&   $ 3.52 \pm 0.07$ \\
%                 &                 &             \\
%$V-M_V$          &$ 7.57  \pm 0.06 $& $ 7.57  \pm 0.07 $ \\
$V_0-M_V$         &$ 7.37  \pm 0.09 $& $  7.37  \pm 0.10 $ \\
Distance \, (pc) &$ 298   \pm  12  $& $ 298   \pm  13  $ \\
 & & \\ 
Abundance:                    &                   &   \\              
\feh\       & \multicolumn{2}{c}   {$-0.12 \pm 0.07$} \\
\noalign{\smallskip}            
\hline
\noalign{\smallskip}
%\multicolumn{3}{l}{\dag\ Not corrected for interstellar absorption/reddening.}
\end{tabular}            
\begin{list}{}{}
\item[$^{\mathrm{a}}$] Observed rotational velocity
\item[$^{\mathrm{b}}$] Equatorial velocity for synchronous rotation
\item[$^{\mathrm{c}}$] Equatorial velocity for pseudo-synchronous rotation
\item[$^{\mathrm{d}}$] Refers to periastron velocity
\item[$^{\mathrm{e}}$] Not corrected for interstellar absorption/reddening
\end{list}
\end{center}
\textsc{Note:}
Bolometric corrections ($BC$) by Flower (\cite{flower96}) have been
assumed, together with
$T_{eff\sun} = 5780$ K, $BC_{\sun} = -0.08$, and $M_{bol\sun} = 4.74$.
\end{minipage}
\end{table}

\section{Absolute dimensions}
\label{sec:absdim}

Absolute dimensions for \BK\ are presented in Table~\ref{tab:bkpeg_absdim}, 
as calculated from the photometric and spectroscopic elements given in 
Tables~\ref{tab:bkpeg_phel} and \ref{tab:bkpeg_spel}.
As seen, masses and radii precise to 0.4--0.5\% and 0.4--1.2\%, respectively,
have been established for the binary components. 

The $V$ magnitudes and $uvby$ indices for the components, included in
Table~\ref{tab:bkpeg_absdim}, were calculated from the combined 
magnitude and indices of the system outside eclipses 
(Table~\ref{tab:bkpeg_std}) and the luminosity ratios between the 
components (Table~\ref{tab:bkpeg_phel}). 
Within errors, the $V$ magnitude and the $uvby$ 
indices of the primary component agree with those measured at central 
secondary eclipse where 98\% ($y$) of the light from the secondary component 
eclipsed (cf. Table~\ref{tab:bkpeg_std}).  

The interstellar reddening $E(b-y) = 0.044 \pm 0.015$, also given in
Table~\ref{tab:bkpeg_absdim}, was determined from the calibration by 
Olsen (\cite{olsen88}), using the $uvby\beta$ standard photometry for the 
combined light outside eclipses.
Within errors, the same reddening is obtained from the indices observed
during the central part of secondary eclipse.
The new intrinsic-colour calibration by Karatas \& Schuster (\cite{ks10})
leads to $E(b-y) = 0.034$.
The model by Hakkila et al. (\cite{hak97}) yields a higher reddening 
of $E(B-V) = 0.14$ or $E(b-y) = 0.10$ in
the direction of and at the distance of \BK, whereas the maps by
Burstein \& Heiles (\cite{bh82}) and Schlegel et al. (\cite{sch98}) give
{\em total} $E(B-V)$ reddenings of 0.07 and 0.05, respectively.

From the individual indices and the calibration by Holmberg et al. 
(\cite{holmberg07}), we derive effective temperatures of
$6265 \pm 85$ and $6285 \pm 90$~K for the primary and secondary
component, respectively, assuming the final \feh\ abundance. 
The temperature uncertainties include those of the $uvby$ indices, 
$E(b-y)$, \feh, and the calibration itself.
Compared to this, the empirical flux scale by Popper (\cite{dmp80}) 
and the $y$ flux ratio between the components (Table~\ref{tab:bkpeg_phel})
yield a well established temperature difference between the components of
$55 \pm 5$~K (excluding possible errors of the scale itself).
Consequently, we assign temperatures of 6265 and 6320~K, which agree with
the corrected spectroscopic results, 6325 and 6345~K, within errors (Sect.\ref{sec:abund}).
As seen from Table~\ref{tab:teff} (combined light), temperatures from the 
$(b-y)$, $c_1$ calibration by Alonso et al. (\cite{alonso96}) and the 
$b-y$ calibration by Ram\'\i rez \& Mel\'endez (\cite{rm05}) are in perfect 
agreement with that from the Holmberg et al. calibration.

The measured rotational velocity for the secondary component
is in perfect agreement with (pseudo)synchronous rotation, whereas
the primary component seems to rotate at a slightly lower rate.

The distance to \BK\ was calculated from the "classical" relation (see e.g. CTB08),
adopting the solar values and bolometric corrections given in
Table~\ref{tab:bkpeg_absdim}, and $A_V/E(b-y) = 4.28$ (Crawford \& Mandwewala \cite{cw76}).
As seen, identical values are obtained for the two components, and 
the distance has been established to 4\%, accounting for all
error sources.
Nearly the same distance is obtained if other $BC$ scales, e.g.
Code et al. (\cite{code76}), Bessell et al. (\cite{bessell98}) and
Girardi et al. (\cite{girardi02}), are adopted.
Also, the empirical $K$ surface brightness - $T_{\rm eff}$ relation
by Kervella et al. (\cite{kervella04}) leads to an identical and perhaps
even more precise (2\%) distance. 
We refer to Clausen (\cite{jvc04}) and Southworth et al. (\cite{southworth05}) 
for details on the use of eclipsing binaries as standard candles.

\section{Comparison with stellar models}
\label{sec:models}

\begin{table}
\caption[]{\label{tab:bkpeg_models}
%Models information and average ages inferred from the observed masses and radii. 
Model parameters and average ages for \BK\ inferred from the observed masses and radii. 
%see Figs.~\ref{fig:bkpeg_mr} and \ref{fig:bkpeg_mr_vrss}.
}
\centering
\begin{tabular}{lccccc} \hline
\hline\noalign{\smallskip}
Grid        & \feh\  &  $Y$  &  $Z$     & Age (Gyr)\\
\noalign{\smallskip}
\hline
\noalign{\smallskip}
Yonsei-Yale     &$-0.190$ & 0.2540& 0.0120    & 2.55\\
                &$-0.120$ & 0.2580& 0.0140    & 2.70\\
                &$-0.105$ & 0.2590& 0.0145    & 2.75\\
                &$-0.050$ & 0.2626& 0.0163    & 2.85\\
\hline
Victoria-Regina &$-0.190$ & 0.2629& 0.0125    & 2.35\\
                &$-0.105$ & 0.2684& 0.0150    & 2.50\\
                &$-0.039$ & 0.2735& 0.0173    & 2.60\\
\hline
\end{tabular}
\end{table}

In the following, we compare the absolute dimensions obtained
for \BK\ with properties of recent theoretical stellar evolutionary models.
We have concentrated on the Yonsei-Yale ($Y^2$) grids by Demarque et al.
(\cite{yale04})\footnote{{\tt http://www.astro.yale.edu/demarque/yystar.html}} 
and the VRSS (scaled-solar abundances of the heavy elements) 
Victoria-Regina grids (VandenBerg et al.,
\cite{vr06})\footnote{{\tt http://www1.cadc-ccda.hia-iha.nrc-cnrc.gc.ca/cvo/
community/VictoriaReginaModels/}} listed in Table~\ref{tab:bkpeg_models}.
The abundance, mass, and age interpolation routines
provided by the $Y^2$ group and the isochrone interpolation routines provided 
by the Victoria-Regina group have been applied.
A summary of the $Y^2$ and VRSS grids and their input physics is 
given by CTB08. Here we just recall the following: The $Y^2$ models include He diffusion,
whereas diffusion processes are not included in the VRSS models.
The $Y^2$ models adopt the enrichment law $Y = 0.23 + 2Z$ 
together with the solar mixture by Grevesse et al. (\cite{gns96}),
and the VRSS models $Y = 0.23544 + 2.2Z$ and the solar mixture by 
Grevesse \& Noels (\cite{gn93}).

\begin{figure}
\epsfxsize=90mm
%\epsfbox{bkpeg_tr_mixed.ps}
\epsfbox{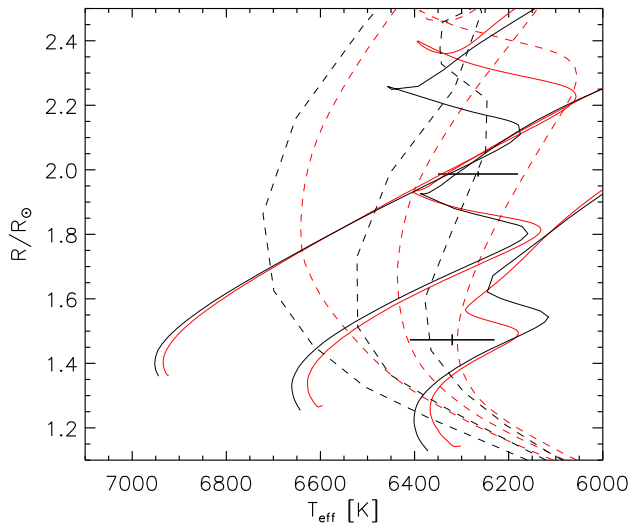}
\caption[]{\label{fig:bkpeg_tr_mixed}
Comparison between $Y^2$ (black) and VRSS (red) models for \feh\,$=-0.105$.
Tracks (solid lines) for 1.2, 1.3, and 1.4 \Msun\ and isochrones (dashed lines)
for 2.0, 2.5, and 3.0 Gyr are included.
The 1.26 and 1.41 \Msun\ components of \BK\ are shown.
}
\end{figure}

\begin{figure}
\epsfxsize=90mm
%\epsfbox{bkpeg_mr_mixed.ps}
\epsfbox{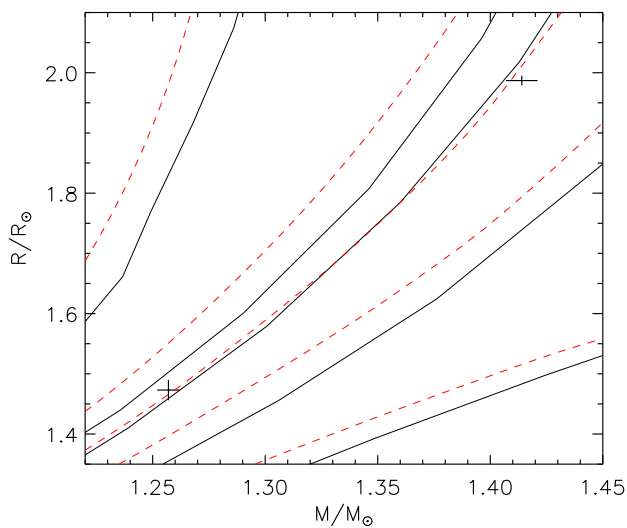}
\caption[]{\label{fig:bkpeg_mr_mixed}
Comparison between $Y^2$ (solid lines, black) and VRSS (dashed lines, red) 
models for \feh\,$=-0.105$.  Isochrones for 1.0, 2.0, 2.5 (VRSS only), 
2.75 ($Y^2$ only), 3.0, and 4.0 Gyr are included.
The 1.26 and 1.41 \Msun\ components of \BK\ are shown.
}
\end{figure}

Both grids include core overshoot. 
For the $Y^2$ models, $\Lambda_{OS} = \lambda_{\rm ov}/H_{\rm p}$ depends on mass and also 
takes into account the composition dependence of $M_{\rm crit}^{\rm conv}$ 
\footnote{Defined as "the mass above which stars continue to have a substantial convective 
core even after the end of the pre-MS phase."}.
The ramping of $\Lambda_{OS} = \lambda_{\rm ov}/H_{\rm p}$ from 
0 at $M_{\rm crit}^{\rm conv}$ to 0.2 at 1.2 $\times$ $M_{\rm crit}^{\rm conv}$ and higher masses 
is done in steps of 0.05.  For the metallicity of \BK, $M_{\rm crit}^{\rm conv}$ is 1.2 \Msun.
The Victoria-Regina group adopts a somewhat different physically based description 
of the core overshoot with the free parameter $F_{\rm over}$ depending on mass 
and metallicity and calibrated observationally via cluster CMDs. 
For the metallicity of \BK, core convection sets in around 1.15 \Msun, and $F_{\rm over}$ 
gradually increases to 0.55 at 1.70 \Msun\ and is kept at this value 
for higher masses. 
Thus, for both grids, the 1.26 and 1.41 \Msun\ components of \BK\ are in their 
ramping regions.

\begin{figure}
\epsfxsize=90mm
%\epsfbox{bkpeg_tr_01398.ps}
\epsfbox{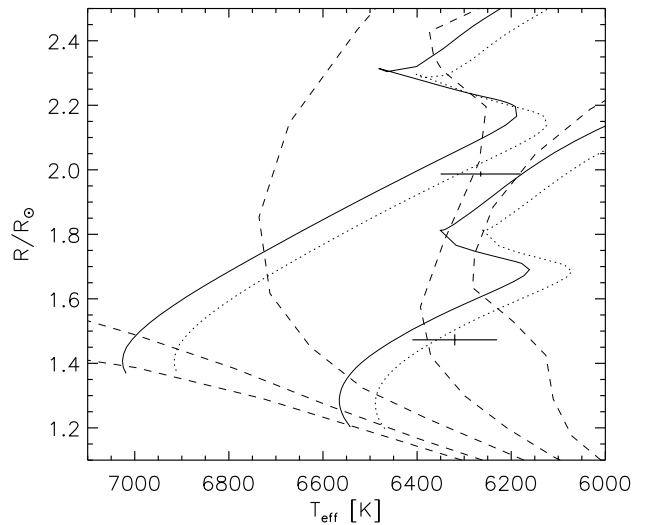}
\caption[]{\label{fig:bkpeg_tr}
\BK\ compared with $Y^2$ models for \feh\,$=-0.12$.
Tracks for the component masses (solid lines) and isochrones for 0.5 and
1.0--4.0 Gyr (dashed; step 1.0 Gyr) are shown.
Tracks (dotted) for \feh\,$=-0.05$ are included.
}
\end{figure}

In Fig.~\ref{fig:bkpeg_tr_mixed}, we compare $Y^2$ and VRSS main-sequence 
evolutionary tracks and isochrones for \feh\,$=-0.105$\footnote{VRSS models
for the observed \feh\,$=-0.12$ are not available}
in the mass region of \BK. 
For 1.3 \Msun\ the tracks nearly coincide, whereas the TAMS positions differ 
for both lower and higher masses, probably related to the individual core overshoot
recipes. In addition, somewhat different ages are predicted along the main-sequence
tracks. This is also illustrated by the isochrones in the mass-radius
diagram shown in Fig.~\ref{fig:bkpeg_mr_mixed}, which represents the most direct
comparison with \BK, since the observed masses and radii are scale independent.
As seen, similar ages of 2.75 ($Y^2$) and 2.50 (VRSS) Gyr are 
predicted for
the components, with a slight preference for the shape of the VRSS
isochrone. Within the observed metal abundance range, 
\feh\,$=-0.12 \pm 0.07$, 
the best fitting isochrones reproduce \BK\ nearly equally well; the 
corresponding average ages are listed in Table~\ref{tab:bkpeg_models}.

\begin{table*}
\caption[]{\label{tab:systems}
Masses, radii, and abundances from Torres et al. (\cite{tag09}) for a subset of 
well-studied binaries with both components in the 1.15--1.70 \Msun\ interval; 
see the text for details.
}
\begin{center}
\begin{tabular}{llrrrrrrrrr} \hline
\hline\noalign{\smallskip}
System    &Colour     &$M_{pri}$& $R_{pri}$ & $M_{sec}$ &  $R_{sec}$&   \feh\   & \feh\ & Age (Gyr)    & \feh\ & Age (Gyr)\\
          &Figs.~\ref{fig:mr_systems}, \ref{fig:mr_systems_vrss}& (\Msun) & (\Rsun)   &  (\Msun)  &  (\Rsun)  & observed  & $Y^2$ &$Y^2$& VRSS   & VRSS \\
\noalign{\smallskip}
\hline
\noalign{\smallskip}
GX\,Gem   &cyan      &  1.488  &    2.326  &    1.467  &    2.236  &          & 0.00  & 2.79& 0.000  & 2.55\vspace{-0.8mm}\\
          &          & $\pm11$ &   $\pm12$ &   $\pm10$ &   $\pm12$ &          &       &     &        &      \\
BK\,Peg   &black     &  1.414  &    1.987  &    1.257  &    1.473  &   $-0.12$&$-0.12$& 2.70&$-0.105$& 2.50\vspace{-0.8mm}\\
          &          & $\pm 7$ &   $\pm 8$ &   $\pm 5$ &   $\pm17$ &    $\pm7$&       &     &        &      \\
BW\,Aqr   &magenta   &  1.377  &    1.786  &    1.479  &    2.062  &   $-0.07$&$-0.08$& 2.43&$-0.105$& 2.15\vspace{-0.8mm}\\
          &          & $\pm21$ &   $\pm43$ &   $\pm19$ &   $\pm44$ &   $\pm11$&       &     &        &      \\
V442\,Cyg &red       &  1.560  &    2.073  &    1.407  &    1.663  &          &$-0.10$& 1.77&$-0.105$& 1.65\vspace{-0.8mm}\\
          &          & $\pm24$ &   $\pm34$ &   $\pm23$ &   $\pm33$ &          &       &     &        &     \\
%         &dashed    & $\pm24$ &   $\pm34$ &   $\pm23$ &   $\pm33$ &          &$ 0.00$& 1.83&$ 0.00$ & 1.67\\
AD\,Boo   &green     &  1.414  &    1.612  &    1.209  &    1.216  &   $+0.10$& +0.10 & 1.77& +0.136 & 1.50\vspace{-0.8mm}\\
          &          & $\pm 9$ &   $\pm14$ &   $\pm 6$ &   $\pm10$ &   $\pm15$&       &     &        &      \\
FS\,Mon   &orange    &  1.632  &    2.052  &    1.462  &    1.629  &          & +0.23 & 1.44& +0.226 & 1.23\vspace{-0.8mm}\\
          &          & $\pm10$ &   $\pm12$ &   $\pm10$ &   $\pm10$ &          &       &     &        &     \\
%         &dashed    & $\pm10$ &   $\pm12$ &   $\pm10$ &   $\pm10$ &           &  0.00 & 1.38& +0.00  & 1.27\\
VZ\,Hya   &brown     &  1.271  &    1.314  &    1.146  &    1.113  &   $-0.20$&$-0.20$& 1.20&$-0.190$& 0.90\vspace{-0.8mm}\\
          &          & $\pm 9$ &   $\pm 5$ &   $\pm 6$ &   $\pm 7$ &   $\pm12$&       &     &        &      \\
V570\,Per &blue      &  1.447  &    1.521  &    1.347  &    1.386  &   $+0.02$& +0.02 & 1.00&  0.000 & 0.75\vspace{-0.8mm}\\
          &          & $\pm 9$ &   $\pm34$ &   $\pm 8$ &   $\pm19$ &    $\pm3$&       &     &        &      \\
HD71636   &gray      &  1.514  &    1.571  &    1.288  &    1.363  &          &  0.00 & 0.80/1.70&  0.000 & 0.70/1.30\vspace{-0.8mm}\\
          &          & $\pm 7$ &   $\pm26$ &   $\pm 6$ &   $\pm26$ &          &       &     &        &      \\
%CD Tau        1.442 0.016 1.798 0.015 1.368 0.016 1.584 0.018 +0.08 0.15 2.18 cyan     0.000  0.136
%V1130 Tau     1.306 0.008 1.489 0.010 1.392 0.008 1.782 0.011 -0.25 0.10 2.13 blue    -0.190 -0.290
\noalign{\smallskip}
\hline
\end{tabular}
\end{center}
\end{table*}

 \begin{figure*}
 \epsfxsize=180mm
%\epsfbox{tr_systems_new.ps}
 \epsfbox{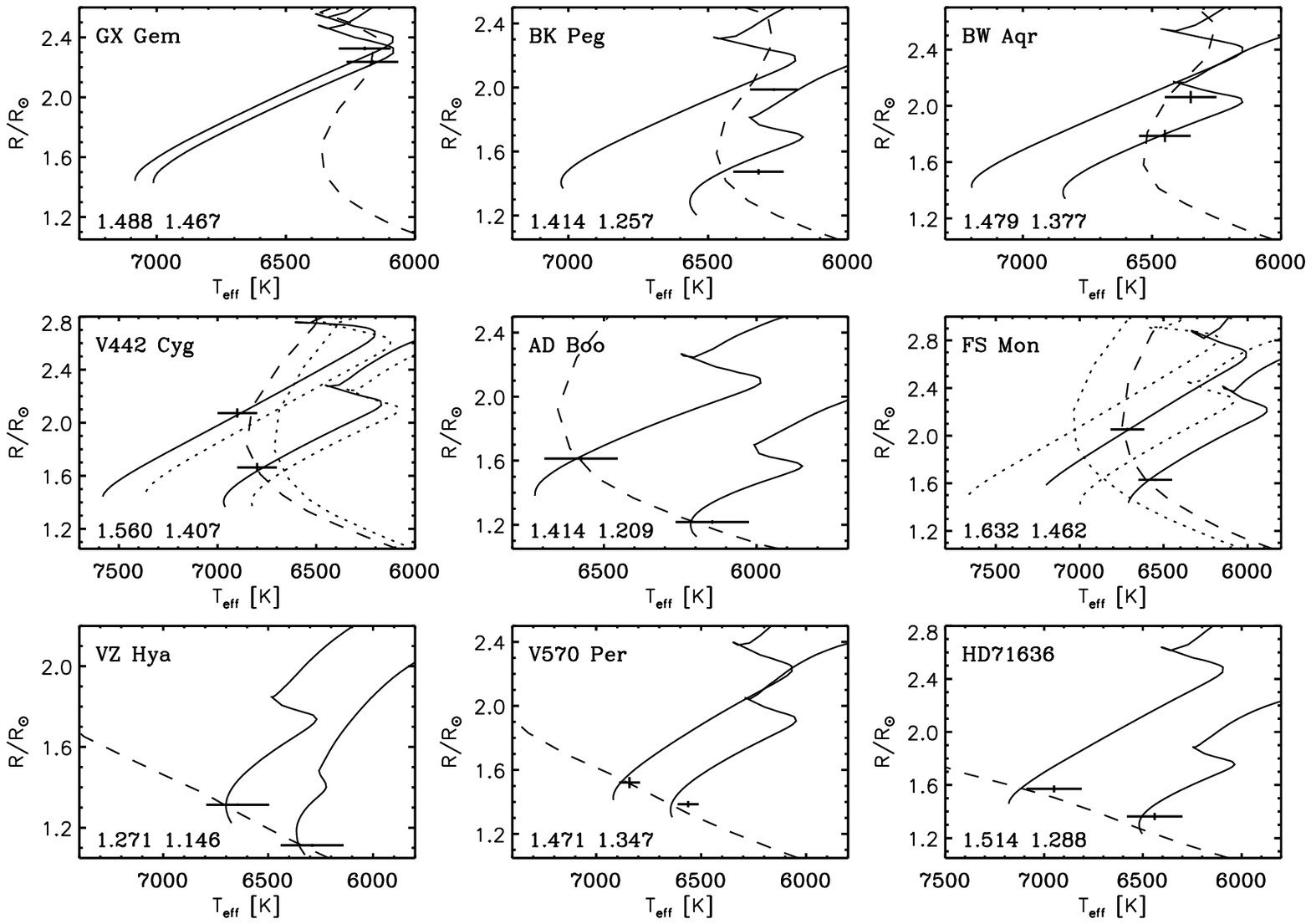}
 \caption[]{\label{fig:tr_systems}
 $Y^2$ evolutionary tracks (full drawn lines) for the binaries included in 
Table~\ref{tab:systems}. The isochrones (dashed lines) correspond to the
average ages inferred from the masses and radii for the \feh\ listed in the
table.
For HD~71636, the 0.80 Gyr isochrone is plotted.
For V442\,Cyg and FS\,Mon, the dotted tracks and isochrones (1.83 and 1.38 Gyr, respectively)
represent solar abundance models.
Component masses (in units of \Msun) are given in the lower left corners.
}
 \end{figure*}

$Y^2$ tracks for the observed masses and abundance of \BK\ are shown in 
Fig.~\ref{fig:bkpeg_tr}. They are, like the VRSS tracks, slightly hotter.
We also recall that the temperatures derived as part of the spectroscopic
abundance analysis are slightly higher than those adopted from the $uvby$
indices (Table~\ref{tab:bkpeg_absdim}).

In conclusion, the $Y^2$ and Victoria-Regina VRSS models, including
their core overshoot treatment and ramping recipes, are able to reproduce
the observed properties of \BK. 
We note, however, that both grids predict a slightly lower age for the 
1.41 \Msun\ primary component than for the 1.26 \Msun\ secondary.
In the following section, we will check whether this tendency is seen for other
well-studied binaries with similar component masses. 

\section{Comparison with other binaries}
\label{sec:binaries}

The review on accurate masses and radii of normal stars by Torres et al. (\cite{tag09}) 
lists 20 binaries\footnote{The 20 binaries can now be supplemented by \BK\ and \object{V1130\,Tau} (Clausen et al. \cite{cohc10}).}
with both components in the 1.15--1.70 \Msun\ interval where
core overshoot is ramped up in the $Y^2$ and Victoria-Regina models.
Here, we will concentrate on the subset listed in Table~\ref{tab:systems}, where
most of the nearly equal-mass binaries have been excluded.
Eight binaries have been selected because the masses of their components
differ significantly; we have adopted a lower mass difference limit of 0.10 \Msun.
In addition, we have included GX\,Gem, since its components have evolved close to the TAMS region.
As mentioned in Sect.~\ref{sec:intro}, Lacy et al. (\cite{ltc08}) found that for 
GX\,Gem, the lowest $\alpha_{\rm ov}$ consistent with observations is approximately
0.18. \feh\ has been measured for five of the systems; we refer to Appendix~\ref{sec:bwaqr} for
details on the abundance analysis of BW\,Aqr.

Fig.~\ref{fig:tr_systems} shows the nine binaries together with $Y^2$ evolutionary
tracks for the measured masses.
For GX\,Gem, \object{V442\,Cyg}, \object{FS\,Mon}, and \object{HD~71636}, \feh\ has
not been measured. The values listed in Table~\ref{tab:systems} were adopted only because
they lead to reasonable model fits for the observed effective temperatures. 
The isochrones correspond to the average age inferred from the measured masses and radii.
To illustrate the abundance dependence, evolutionary tracks and isochrones for solar
metallicity have been included for V442\,Cyg and FS\,Mon. 
As seen, the nine binaries cover the main-sequence almost from the ZAMS to the TAMS.

\begin{figure}
\epsfxsize=90mm
%\epsfbox{mr_systems_new.ps}
\epsfbox{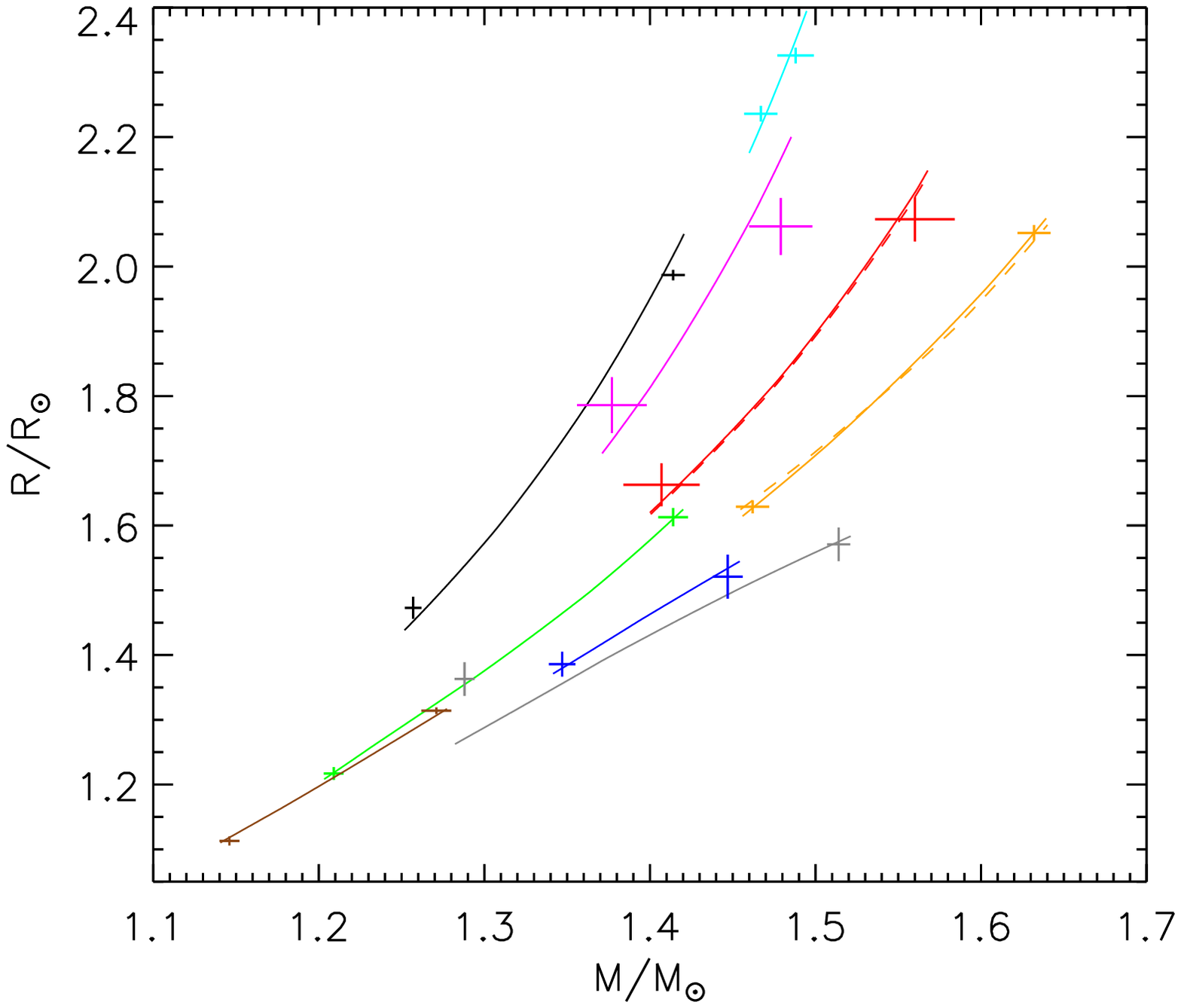}
\caption[]{\label{fig:mr_systems}
Comparison between $Y^2$ isochrones (full drawn lines) and the binaries 
listed in Table~\ref{tab:systems}; we refer to the table for colour codes, \feh\ and ages.
For V442\,Cyg and FS\,Mon, the dashed lines are solar abundance isochrones for
1.83 and 1.38 Gyr, respectively.
For HD~71636, the 0.80 Gyr isochrone is plotted.
}
\end{figure}

\begin{figure}
\epsfxsize=90mm
%\epsfbox{mr_systems_new_vrss.ps}
\epsfbox{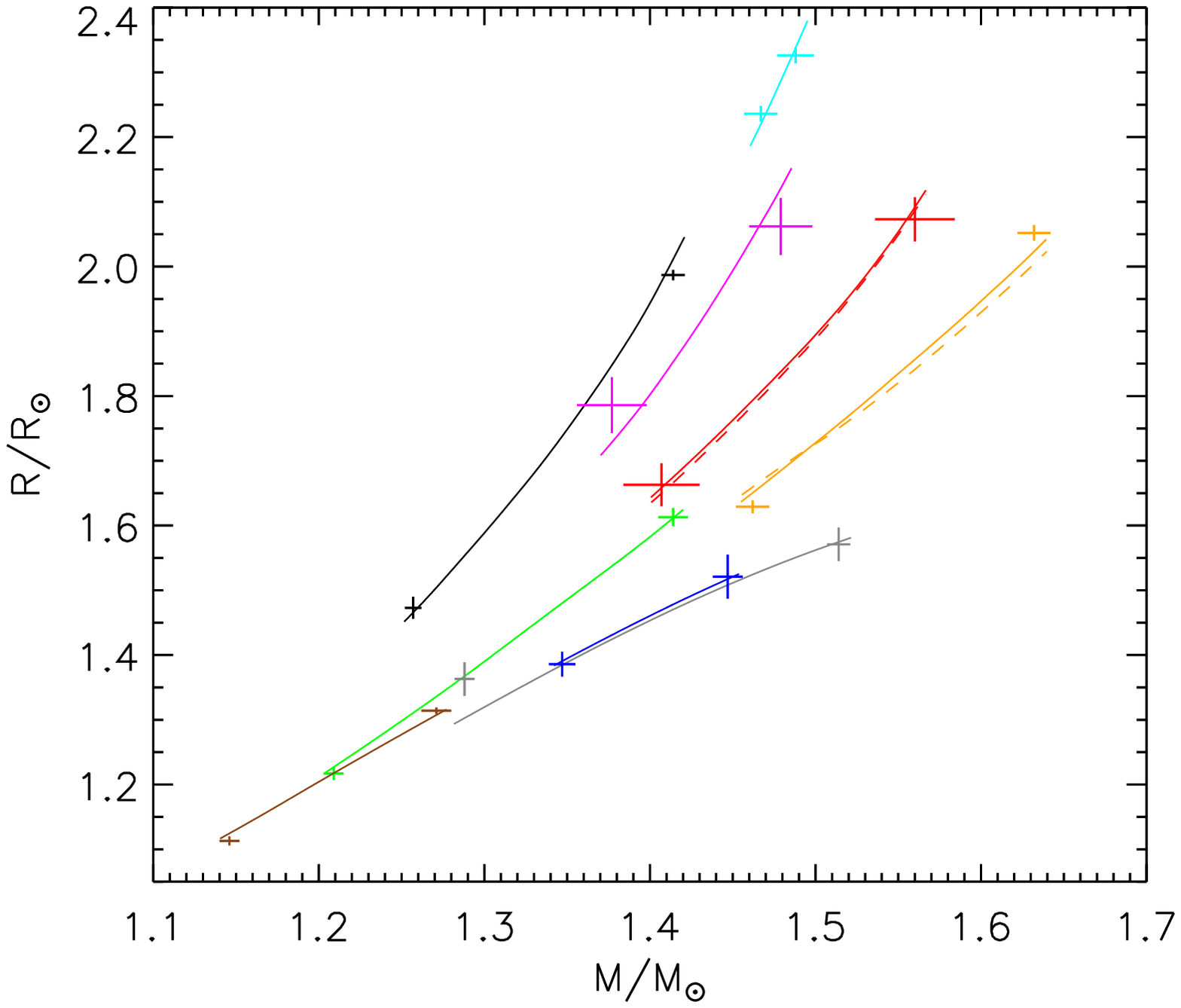}
\caption[]{\label{fig:mr_systems_vrss}
Comparison between VRSS isochrones (full drawn lines) and the binaries 
listed in Table~\ref{tab:systems}; we refer to the table for colour codes, \feh\ and ages.
For V442\,Cyg and FS\,Mon, the dashed lines are solar abundance isochrones for
1.67 and 1.27 Gyr, respectively.
For HD~71636, the 0.70 Gyr isochrone is plotted.
}
\end{figure}

In Figs.~\ref{fig:mr_systems} ($Y^2$) and \ref{fig:mr_systems_vrss} (VRSS), we concentrate 
on comparisons based on the scale-independent masses and radii. Since abundance interpolation
software is not available for the VRSS models, we have used those with \feh\ closest to
the measured values.
The average ages are listed in Table~\ref{tab:systems}, and we note the following:
\begin{itemize}
\item
The VRSS models predict lower ages than the $Y^2$ models for all systems.
\item
For the four binaries with components in the upper half of the main-sequence band
(GX\,Gem, BK\,Peg, BW\,Aqr, V442\,Cyg), both grids tend to predict lower ages for 
the more massive component than for the less massive one, although the difference is
certainly very marginal for the nearly equal-mass system GX\,Gem. 
These systems span the 1.26--1.56 \Msun\ interval.
\item
The $Y^2$ isochrones fit FS\,Mon (1.63 + 1.46 \Msun) better than the VRSS isochrones,
which predict a higher age for the more massive component than for the less
massive one.
\item
Both grids fit the less evolved binaries \object{AD\,Boo} and \object{VZ\,Hya} quite well,
perhaps with a slight preference for $Y^2$.
\item
For the little evolved binary \object{V570\,Per}, VRSS is marginally better than $Y^2$.
\item
HD~71636 is not fitted well by any of the grids. The very different ages predicted 
for the components are listed in Table~\ref{tab:systems}.  We believe,
however, that this is due to problems with the published radii. A re-analysis
of the light curves by Henry et al. (\cite{hfsg06}) indicates that the ratio
between the (relative) radii is significantly more uncertain than given by the
authors\footnote{We have used the larger errors listed by Torres et al. (\cite{tag09}), 
which also seem to be underestimated.},
most probably because the secondary eclipse is far from being well-covered.
\end{itemize}

It is outside the scope of this paper to propose specific modifications of 
the model physics, but we believe that this sample of binaries (except HD~71636)
can be used to fine-tune the core overshoot treatment in terms of mass and metal
abundance. It might also be relevant to include investigations of other model 
ingredients, e.g. diffusion processes and the adopted helium-to-metal enrichment 
ratio, and to clarify why the $Y^2$ models predict higher ages than the VRSS models. 
In order to avoid possible interpolation errors, specific models 
for the observed masses and metal abundances should be calculated and
small age steps applied.
On the observational side, spectroscopic metal abundances should
be determined for the four binaries lacking this information, and it should
perhaps be considered to re-observe BW\,Aqr, V442\,Cyg, and HD~71636 to 
improve their masses and/or radii.

Also, we suggest to supplement the sample by the unique K0IV+F7V binary 
AI~Phe ($1.23+1.19$ \Msun), recently re-discussed by Torres et al. (\cite{tag09}),
as well as by a number of new F-type systems we are presently studying. 
In addition, it would be highly relevant to include F-type binary members of
open clusters. 

\section{Summary and conclusions}
\label{sec:sum}

From state-of-the-art observations and analyses, precise (0.4--1.2\%) absolute
dimensions have been established for the components of the late F-type
detached eclipsing binary \BK\ ($P = 5\fd49$, $e$ = 0.0053);
see Table~\ref{tab:bkpeg_absdim}.
A detailed spectroscopic analysis yields an iron abundance relative to the Sun
of \feh\,$=-0.12\pm0.07$ and similar relative abundances for Si, Ca, Sc, Ti, Cr,
and Ni.
The measured rotational velocities are $16.6 \pm 0.2$ (primary) and
$13.4 \pm 0.2$ (secondary) \kms. For the secondary component this corresponds
to (pseudo)synchronous rotation, whereas the primary component seems to
rotate at a slightly lower rate. 

The 1.41 and 1.26 \Msun\ components of \BK\ have evolved to the upper half 
of the main-sequence band. 
Yonsei-Yale and Victoria-Regina solar scaled evolutionary models for the
observed metal abundance reproduce \BK\ at ages of 2.75 and 2.50 Gyr, 
respectively, but tend to predict a lower age for the more massive primary 
component than for the secondary.
If real, this might be due to less than perfect calibration of the amount of
convective core overshoot of the models as function of mass 
(and metal abundance).

For this reason, we have performed model comparisons for a sample of eight 
additional well-studied binaries with component masses in the 1.15--1.70 
\Msun\ interval where convective core overshoot is gradually ramped up in
the models; see Table~\ref{tab:systems}. 
We find that {\it a}) the Yonsei-Yale models systematically predict higher ages 
than the Victoria-Regina models, and that {\it b}) the three other most evolved 
systems in the sample share the age difference trend seen for \BK.

We propose to use the sample to fine-tune the core overshoot treatment, 
as well as other model ingredients, and to clarify why the two model grids 
predict different ages. The sample should be expanded by a number of new 
F-type systems under study, binary cluster members, and the unique 
K0IV+F7V binary AI\,Phe (1.23+1.19 \Msun).

\begin{acknowledgements}
It is a great pleasure to thank the many colleagues and students,
who have shown interest in our project and have participated in the
extensive (semi)automatic observations of \BK\ at the SAT:
Sylvain Bouley,
Christian Coutures,
Thomas H. Dall,
Mathias P. Egholm,
Pascal Fouque,
Lisbeth F. Grove,
Anders Johansen,
Erling Johnsen,
Bjarne R. J{\o}rgensen,
Bo Milvang-Jensen,
Alain Maury,
John D. Pritchard, and
Samuel Regandell.
Excellent technical support was received from the staffs of Copenhagen
University and ESO, La Silla.
We thank J.~M. Kreiner for providing a complete list of published times 
of eclipse for \BK.
A. Kaufer, O. Stahl, S. Tubbesing, and B. Wolf kindly obtained
the two FEROS spectra of BW\,Aqr during Heidelberg/Copenhagen guaranteed
time in 1999.

The projects "Stellar structure and evolution -- new challenges from
ground and space observations" and "Stars: Central engines of the evolution
of the Universe", carried out at Copenhagen University and Aarhus University,
are supported by the Danish National Science Research Council.

The following internet-based resources were used in research for
this paper: the NASA Astrophysics Data System; the SIMBAD database
and the VizieR service operated by CDS, Strasbourg, France; the
ar$\chi$iv scientific paper preprint service operated by Cornell University;
the VALD database made available through the Institute of Astronomy,
Vienna, Austria; the MARCS stellar model atmosphere library.
This publication makes use of data products from the Two Micron
All Sky Survey, which is a joint project of the University of
Massachusetts and the Infrared Processing and Analysis Center/California 
Institute of Technology, funded by the National
Aeronautics and Space Administration and the National Science
Foundation.
\end{acknowledgements}

{}

\begin{appendix}
\section{Chemical abundances for BW\,Aqr}
\label{sec:bwaqr}

Absolute dimensions for the late F-type eclipsing binary BW\,Aqr were
published by Clausen (\cite{jvc91}), who mentioned that the $uvby\beta$
photometry indicates a metallicity slightly above solar. 
We note that the primary component (star A in Clausen \cite{jvc91}), 
eclipsed at the deeper minimum at phase 0.0, is less massive, smaller, but hotter than
the secondary component.

Here we present the results from an abundance
analysis based on two high-resolution spectra observed with the FEROS
fibre echelle spectrograph at ESO, La Silla in August 1999; see Table~\ref{tab:feros_bwaqr}.
Details on the spectrograph, the reduction of the spectra,
and the basic approach followed in the abundance analysis are
described by CTB08.
As for \BK, we have used the VWA tool for the abundance analysis,
and we refer to Sect.~\ref{sec:abund} for further information on atmosphere
models, atomic data information, $\log(gf$) adjustments etc. 
One important difference is, however, that for BW\,Aqr disentangling 
is not possible, and the analysis is therefore based on double-lined spectra.

The effective temperatures, surface gravities, rotational velocities, and
microturbulence velocities listed in Table~\ref{tab:bwaqr_param} were adopted.
The temperatures were determined by requiring that Fe \ione\ abundances were
independent of line excitation potentials. 
They are slightly lower than 
determined by Clausen (\cite{jvc91}), who obtained $6450 \pm 100$ K (primary)
and $6350 \pm 100$ K (secondary). However, new and better (unpublished) $uvby$ 
photometry for BW\,Aqr and the calibration by Holmberg et al. (\cite{holmberg07}) lead
to 100 K lower values, assuming a reddening of $E(b-y) = 0.03$.
Microturbulence velocities were tuned until Fe \ione\ abundances are
independent of line equivalent widths. 

The abundances derived from all useful lines in both spectra
are presented in Table~\ref{tab:bwaqr_abund}.
The equivalent widths measured in the two double-lined spectra are listed
in Tables~16 (primary) and 17 (secondary), which will only be available in
electronic form.
Comparing the results from the two spectra, we find that they agree within
0.05 dex. 
We have only included lines with {\em measured} equivalent
widths above 10 m{\AA} and below 45 m{\AA} (primary) and 55 m{\AA} (secondary).
The lines are diluted by factors of about 2.2 (primary)
and 1.8 (secondary), meaning that lines with {\em intrinsic} strengths
above 100 m{\AA} are excluded.
The \feh\ results for the two components differ by 0.06 dex, but 
for each component the results from Fe\,\ione\ and Fe\,\itwo\ lines agree
well. The mean value for all measured Fe lines is \feh\,=$-0.07 \pm 0.01$
(rms of mean).

\begin{table}
\caption[]{\label{tab:feros_bwaqr}
Log of the FEROS observations of BW\,Aqr.
%jvc080827
}
\begin{minipage}{\columnwidth}
\centering
\renewcommand{\footnoterule}{}  % to avoid a line before footnotes
\begin{tabular}{ccrrr}
\hline
\hline\noalign{\smallskip}
HJD$-$2\,400\,000\footnote{Refers to mid-exposure}  & phase &t$_{exp}$\footnote{Exposure time in seconds} & S/N\footnote{Signal-to-noise ratio measured around 6070 {\AA}}\\
\noalign{\smallskip}
\hline
\noalign{\smallskip}
51393.82092& 0.1290 & 3600 & 140 \\ %    f8942  
51394.70920& 0.2612 & 3600 & 200 \\ %    f9002  
\hline
\end{tabular}
\end{minipage}
\end{table}

\begin{table}   
\caption[]{\label{tab:bwaqr_param}
Astrophysical data adopted for the abundance analysis of BW\,Aqr.
}
\begin{center}    
\begin{tabular}{lrr} \hline    
\noalign{\smallskip}    
\hline    
\noalign{\smallskip}    
                     &    Primary       &    Secondary      \\ 
\noalign{\smallskip}    
\hline    
\noalign{\smallskip}    
$T_{\mbox{\scriptsize eff}}\,$ (K)  &   6300     &   6225 \\
$\log g$ (cgs)                      &  4.075     &   3.981 \\
$v \sin i$ (\kms)                   &  14.0      &  14.0    \\  
$v_{micro}$ (\kms)                  &   1.70     &   1.50   \\  
\noalign{\smallskip}            
\hline
\noalign{\smallskip}
\end{tabular}            
\end{center}
\end{table}

\begin{table}
\caption[]{\label{tab:bwaqr_abund}
Abundances ($[\mathrm{El./H}]$) for the primary and secondary
components of BW\,Aqr determined from the two FEROS spectra.
}
\begin{center}
\begin{tabular}{lrlrrlr} \hline
\hline\noalign{\smallskip}
             &  \multicolumn{3}{c}{Primary} & \multicolumn{3}{c}{Secondary} \\
Ion          &[El./H]&  rms& N$_t$/N$_l$&[El./H]& rms & N$_t$/N$_l$  \\
\noalign{\smallskip}
\hline
 Si\ione\     &$-0.12$& 0.02&  3/2  &$-0.05$& 0.12& 13/7 \\  %alpha   
 Ca\ione\     &$-0.18$& 0.13&  3/2  &$-0.00$& 0.09&  7/4  \\  %alpha    
 Sc\itwo\     &       &     &       &$-0.08$& 0.08&  3/2  \\            
 Ti\ione\     &$-0.02$& 0.09&  3/2  &       &     &      \\  %alpha   
 Cr\ione\     &$-0.12$& 0.17&  4/2  &$-0.04$& 0.12&  5/4\\  %          
 Fe\ione\     &$-0.11$& 0.16& 65/41 &$-0.05$& 0.11& 80/50   \\
 Fe\itwo\     &$-0.09$& 0.18& 14/8  &$-0.03$& 0.11& 14/8 \\
 Ni\ione\     &$-0.05$& 0.14& 18/11 &$-0.12$& 0.10& 15/10\\    %      
\noalign{\smallskip}
\hline
\end{tabular}            
\end{center}
\textsc{Note:}
N$_t$ is the total number of lines used per ion, and N$_l$ is the
number of different lines used per ion.
Ions with at least 3 lines measured are included.
\end{table}

Changing the model temperatures by $\pm 100$~K modifies \feh\
from the Fe\,\ione\ lines by about $\pm0.10$ dex, whereas almost no
effect is seen for the Fe\,\itwo\ lines.
If 0.30 \kms\ higher microturbulence velocities are adopted,
\feh\ decreases by about 0.05 dex for both neutral and ionized lines.
Taking these contributions to the uncertainties into account,
we adopt \feh\,$=-0.07\pm0.11$ for BW\,Aqr.
Similar abundances are obtained for the few other ions listed in
Table~\ref{tab:bwaqr_abund}.

\end{appendix}

\listofobjects
\end{document}